\documentclass[11pt]{article}
\usepackage{graphicx}
\usepackage{amsmath,amssymb}
\usepackage{txfonts,bm}
\usepackage{abstract}
\usepackage[title]{appendix}
\usepackage{macro} 

\allowdisplaybreaks[4]

\graphicspath{{figures/}}

\begin{document}
\begin{center}
\huge{\textsc{Composite Spatial Monte Carlo Integration Based on Generalized Least Squares}} \\
\vspace{5mm}
\large{Kaiji Sekimoto\footnote{k.sekimoto1002@gmail.com} and Muneki Yasuda} \\
\vspace{2mm}
\normalsize{Graduate School of Science and Engineering, Yamagata University, Japan}
\end{center}

\vspace{2mm}
\begin{abstract}
\vspace{-2.5mm}
Although evaluation of the expectations on the Ising model is essential in various applications, 
it is mostly infeasible because of intractable multiple summations. 
Spatial Monte Carlo integration (SMCI) is a sampling-based approximation. 
It can provide high-accuracy estimations for such intractable expectations. 
To evaluate the expectation of a function of variables in a specific region (called target region), 
SMCI considers a larger region containing the target region (called sum region). 
In SMCI, the multiple summation for the variables in the sum region is precisely executed, 
and that in the outer region is evaluated by the sampling approximation such as the standard Monte Carlo integration.
It is guaranteed that the accuracy of the SMCI estimator improves monotonically as the size of the sum region increases.
However, a haphazard expansion of the sum region could cause a combinatorial explosion.
Therefore, we hope to improve the accuracy without such an expansion.
In this paper, based on the theory of generalized least squares (GLS), a new effective method is proposed by combining multiple SMCI estimators. 
The validity of the proposed method is demonstrated theoretically and numerically. 
The results indicate that the proposed method can be effective in the inverse Ising problem (or Boltzmann machine learning).
\end{abstract}

\section{Introduction} \label{sec:Introduction} 

The Ising model, which is also known as Boltzmann machine in the machine-learning field~\cite{Ackley1985,Roudi2009}, is a traditional and important stochastic model in statistical mechanics.  
Some variants of Boltzmann machine, such as restricted Boltzmann machine (RBM)~\cite{smolensky1986,hinton2002} 
and deep Boltzmann machine (DBM)~\cite{salakhutdinov2009}, also exist. 
Boltzmann machine and its variants have been developed mainly in the machine-learning field; however, recently, they are also actively investigated in physics~\cite{decelle2021,chen2018,nomura2021,torlai2018,carleo2017}.
The evaluation of expectations on the Ising model is quite fundamental in almost all applications of this model.
However, this evaluation is infeasible in general because it contains multiple summations over all variables. 

Monte Carlo integration (MCI) is one of the most popular methods for the evaluation of intractable expectations.  
In an Ising model with variables $\bm{x} = \{x_i \mid i \in \mcal{V}\}$ (where $\mcal{V}$ is the set of induces of all variables),
suppose we want to obtain the expectation of function $f_{\mcal{T}}$ of variables in \textit{target region} $\mcal{T} \subseteq \mcal{V}$.  
In MCI,  the expectation is simply evaluated by the sample average of $f_{\mcal{T}}$ over a sample set generated from the Ising model 
using a sampling method, such as Gibbs sampling~\cite{Geman&Geman1984} and parallel tempering~\cite{swendsen1986,hukushima1996}.  
This means that MCI does not fully utilize the structure of the model even though the structure is known.
Recently, a smaller variance MCI, which is a spatial extension of MCI, called \textit{spatial Monte Carlo integration} (SMCI), was proposed~\cite{yasuda2015,yasuda2021}.
SMCI is regarded as the application of Rao-Blackwellization~\cite{liu2001} to Markov random fields.
In SMCI, we define region $\mcal{U}$, which contains the target region as its subregion; particularly, $\mcal{U}$ is called \textit{sum region}.
The expectation of function $f_{\mcal{T}}$ is evaluated as follows: 
multiple summation over variables in the sum region is precisely executed, 
while multiple summation over variables outside of the sum region is evaluated through sampling approximation similar to the standard MCI.
SMCI has proven to be statistically more accurate than MCI, 
and its approximation accuracy improves monotonically as the size of the sum region increases~\cite{yasuda2015,yasuda2021}. 
Although an increase in the size of the sum region leads to an exponential increase in the computational cost, SMCI is practical when it is small. 
SMCI is effective in the inverse Ising problem~\cite{yasuda2015,yasuda2018,yasuda2021,katsumata2021}, which is also known as Boltzmann machine learning. 
Empirically, the SMCI-based learning method is better than some existing learning methods~\cite{yasuda2018,katsumata2021}, such as contrastive divergence~\cite{hinton2002}, maximum pseudo-likelihood~\cite{besag1975,Aurell2012}, 
ratio matching~\cite{hyvarinen2007}, and minimum probability flow~\cite{dickstein2011}. 
The detailed explanation of SMCI is described in section \ref{sec:SMCI}.

Suppose we have two different sum regions, $\mcal{U}_{\mrm{I}}$ and $\mcal{U}_{\mrm{II}}$, for the purpose of the evaluation of $f_{\mcal{T}}$. 
When $\mcal{U}_{\mrm{I}} \subset \mcal{U}_{\mrm{II}}$, we should select the SMCI estimator with $\mcal{U}_{\mrm{II}}$ because it is guaranteed to be more accurate as per the theory of SMCI.
However, when an inclusion relation does not exist between both regions, we cannot determine which SMCI estimator is more accurate. 
A SMCI estimator with a larger sum region containing $\mcal{U}_{\mrm{I}}$ and $\mcal{U}_{\mrm{II}}$ is surely more accurate 
than the two SMCI estimators with $\mcal{U}_{\mrm{I}}$ and $\mcal{U}_{\mrm{II}}$, respectively.
However, a haphazard expansion of the sum region could cause a combinatorial explosion.  
We hope to obtain a more effective estimator without such an expansion. 
In this paper, based on generalized least squares (GLS)~\cite{aitken1936,greene2003,lee2018}, 
we propose a method creating the desired estimator by combining the given SMCI estimators, referred to as \textit{composite spatial Monte Carlo integration} (CSMCI).
Per the theory of GLS, an estimator obtained from CSMCI is guaranteed to be more accurate than its components (i.e., given SMCI estimators); furthermore, it is ensured to be the best unbiased estimator (BUE) (see section \ref{ssec:theoretical}).  
However, unfortunately, CSMCI is not practical in general because it includes the evaluation of an intractable covariance matrix. 
We also propose an approximation of CSMCI for its implementation.

The remainder of this paper is organized as follows.
The Ising model used in this study is defined in section \ref{sec:Ising_model}.
The detailed explanation of SMCI is provided in section \ref{sec:SMCI}. 
In section \ref{sec:CSMCI}, we introduce the proposed method, that is, CSMCI. 
Subsequently, the validity of the estimators obtained from the proposed method is verified 
from both theoretical and experimental perspectives in sections \ref{ssec:theoretical} and \ref{ssec:numerical}, respectively.  
In section \ref{sec:inverse_ising_problem}, we apply the proposed method to the inverse Ising problem, 
and then demonstrate its performance through numerical experiments.
Finally, the summary and future works are described in section \ref{sec:conclusion}. 

\section{Ising Model} \label{sec:Ising_model} 

Consider an undirected graph, $G(\mcal{V},\mcal{E})$, consisting of $n$ vertices, where $\mcal{V} :=\{1,2,\cdots,n\}$ is the set of vertices 
and $\mcal{E}$ is the set of undirected edges. 
The undirected edge between vertices $i$ and $j$ is labeled by $(i,j)$.
Labels $(i,j)$ and $(j,i)$ indicate the same edge. 
On the graph, an energy function (or a Hamiltonian) is defined as
\begin{align}
E(\bm{x};\theta) := -\sum_{i\in\mcal{V}} h_i x_i -\sum_{(i,j)\in\mcal{E}} J_{i,j}x_ix_j,
\label{eq:def_energy_fuction}
\end{align} 
where $\{h_i \}$ and $\{J_{i,j}\}$, which are collectively denoted by $\theta$, are the external fields and exchange interactions, respectively. 
The exchange interactions are symmetric with respect to their induces, that is, $J_{i,j} = J_{j,i}$.
Here, $\bm{x}:=\{x_i\in\mathcal{X}\mid i\in\mathcal{V}\}$ is a set of random variables assigned on the vertices, where $\mcal{X}$ is the sample space of $x_i$.

In the \textit{standard} Ising model, the sample space is defined as $\mcal{X}=\{-1,+1\}$; however, in this paper, the sample space is not restricted to $\mcal{X}=\{-1,+1\}$, and it can be a discrete or continuous sample space.
Therefore, the Ising model referred in this paper includes the \textit{standard} Ising model.
Using the energy function in equation (\ref{eq:def_energy_fuction}), an Ising model is expressed as 
\begin{align}
P_{\theta}(\bm{x}) := \frac{1}{Z(\theta)}\exp \big(-E(\bm{x};\theta) \big), 
\label{eq:def_ising_model}
\end{align}
where $Z(\theta)$ denotes the partition function which is expressed as
\begin{align}
Z(\theta) := \sum_{\bm{x}}\exp(-E(\bm{x};\theta)), 
\label{eq:partition_function}
\end{align}
where $\sum_{\bm{x}} := \sum_{x_1\in\mathcal{X}}\sum_{x_2\in\mathcal{X}}\cdots\sum_{x_n\in\mathcal{X}} = \prod_{i\in\mathcal{V}}\sum_{x_i\in\mathcal{X}}$ is the summation over all possible realizations of $\bm{x}$; when $\mcal{X}$ is a continuous space, the sum is replaced with the corresponding integration.
The Ising model is known as a Boltzmann machine in the machine learning field~\cite{Ackley1985}.
The set of variables in a region $\mcal{A} \subseteq \mcal{V}$ is denoted by $\bm{x}_{\mcal{A}}$ (i.e., 
$\bm{x}_{\mcal{A}} = \{x_i \mid i \in \mcal{A}\}$), 
and the summation over all possible realizations of $\bm{x}_{\mcal{A}}$ is denoted by 
$\sum_{\bm{x}_{\mcal{A}}}:= \prod_{i\in\mathcal{A}}\sum_{x_i\in\mathcal{X}}$.

On the Ising model, consider a function $f$ of the variables in a \textit{target region} $\mcal{T} \subseteq \mcal{V}$.
The expectation of the function is written as:
\begin{align}
\mexp{f(\bm{x}_{\mcal{T}})}
:= \sum_{\bm{x}}f(\bm{x}_{\mcal{T}})P_{\theta}(\bm{x}) 
= \sum_{\bm{x}_{\mcal{T}}}f(\bm{x}_{\mcal{T}})P_{\theta}(\bm{x}_{\mcal{T}}), 
\label{eq:expectation_f(x_T)}
\end{align}
where $P_{\theta}(\bm{x}_{\mcal{T}}) = \sum_{\bm{x}_{\mcal{V}\setminus\mcal{T}}}P_{\theta}(\bm{x})$ is the marginal distribution of the Ising model.
Except for some special cases, the evaluation of this expectation is computationally infeasible because of the multiple summation whose cost exponentially grows with an increase in the size of the model.
SMCI, described in the following section, is a sampling-based approximation that can effectively evaluate such an intractable expectation~\cite{yasuda2015,yasuda2021}.

Equations given in the subsequent sections are formulated for the case where  $\mcal{X}$ is a discrete space. However, those equations can be applied to the case where $\mcal{X}$ is a continuous space by replacing the summations over $\mcal{X}$ with the integrations over $\mcal{X}$.

\section{Spatial Monte Carlo Integration} \label{sec:SMCI} 

\begin{figure}[t]
\centering
\includegraphics[width=0.4\linewidth]{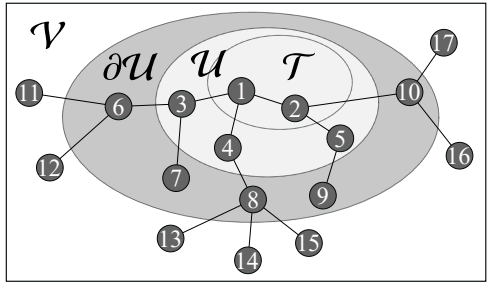}
\caption{Illustration of the three regions for SMCI. 
In this illustration, the target, sum, and sample regions are $\mcal{T} = \{1,2\}$, $\mcal{U} = \{1,2,3,4,5\}$, 
and $\partial\mathcal{U}=\{6,7,8,9,10\}$, respectively.}
\label{fig:region_sheme}
\end{figure}

In this section, we briefly explain SMCI, based on the Ising model introduced in the previous section. 
Suppose that the set of i.i.d. $N$ sample points, 
$\mcal{S} := \big\{ \mbf{s}^{(\mu)} \mid \mu=1,2,\cdots,N \big\}$, drawn from $P_{\theta}(\bm{x})$ is obtained, 
where $\mbf{s}^{(\mu)} := \{ \mrm{s}_i^{(\mu)} \in \mcal{X} \mid i \in \mcal{V}\}$ is the $\mu$th sample point. 
In the standard MCI, using the sample set, the expectation in equation (\ref{eq:expectation_f(x_T)}) is approximated by 
\begin{align}
\mexp{f(\bm{x}_{\mcal{T}})}\approx \sum_{\bm{x}} f(\bm{x}_{\mcal{T}}) Q_{\mcal{S}}(\bm{x}) 
= \frac{1}{N}\sum_{\mu=1}^N f\big(\mathbf{s}_{\mathcal{T}}^{(\mu)}\big),
\label{eq:mci}
\end{align}
where $\mbf{s}_{\mcal{T}}^{(\mu)}$ denotes the $\mu$th sample point on the target region $\mathcal{T}$, i.e., $\mbf{s}_{\mcal{T}}^{(\mu)} := \{ \mrm{s}_i^{(\mu)} \mid i \in \mcal{T}\}$.  
Here, $Q_{\mcal{S}}(\bm{x})$ is the empirical distribution (or the sample distribution) of $\mathcal{S}$, defined by
\begin{align}
Q_{\mcal{S}}(\bm{x}) := \frac{1}{N}\sum_{\mu=1}^N \prod_{i\in\mcal{V}}\delta \big(x_i,\mathrm{s}_i^{(\mu)}\big), 
\label{eq:empir_distribution}
\end{align}
where $\delta$ is the delta function (which is the Kronecker when $\mcal{X}$ is a discrete space, or is the Dirac when $\mcal{X}$ is a continuous space). 

In SMCI, for the target region, region $\mcal{U}$, called the \textit{sum region}, is introduced such that $\mcal{T}\subseteq\mcal{U}\subseteq\mcal{V}$. 
Based on the sum region, the right-hand side of equation (\ref{eq:expectation_f(x_T)}) can be decomposed as 
\begin{align}
\mexp{f(\bm{x}_{\mcal{T}})}&= \sum_{\bm{x}} f(\bm{x}_{\mcal{T}})P_{\theta}(\bm{x}_{\mcal{U}}\mid\bm{x}_{\mcal{V}\setminus\mcal{U}})
P_{\theta}(\bm{x}_{\mcal{V}\setminus\mcal{U}}).
\label{eq:expectation_f(x_T)_trans-0}
\end{align}
From the spatial Markov property of the Ising model, the conditional distribution in this equation is rewritten as 
\begin{align}
P_{\theta}(\bm{x}_{\mcal{U}}\mid\bm{x}_{\mcal{V}\setminus\mcal{U}}) 
= P_{\theta}(\bm{x}_{\mathcal{U}}\mid\bm{x}_{\partial\mcal{U}}), 
\label{eq:conditional_distribution}
\end{align}
where $\partial\mcal{U} := \{j\mid(i,j)\in\mcal{E}, i\in\mcal{U}, j\in \mcal{V} \setminus \mcal{U}\}$ denotes the outer adjacent region of $\mathcal{U}$, which is called the \textit{sample region}. 
The illustration of the three regions (i.e., target, sum, and sample regions) for SMCI is shown in Figure \ref{fig:region_sheme}.
Using the conditional distribution in equation (\ref{eq:conditional_distribution}) 
and the marginalizing operation in equation (\ref{eq:expectation_f(x_T)_trans-0}) yields
\begin{align}
\mexp{f(\bm{x}_{\mcal{T}})}&= \sum_{\bm{x}_{\mcal{U}}}f(\bm{x}_{\mcal{T}})
\sum_{\bm{x}_{\partial \mcal{U}}}P_{\theta}(\bm{x}_{\mathcal{U}}\mid\bm{x}_{\partial\mcal{U}})
P_{\theta}(\bm{x}_{\partial\mcal{U}}).
\label{eq:expectation_f(x_T)_trans-1}
\end{align}
The SMCI estimator for the true expectation is obtained 
by replacing $P_{\theta}(\bm{x}_{\partial\mcal{U}} )$ in equation (\ref{eq:expectation_f(x_T)_trans-1}) with the corresponding marginal distribution of $Q_{\mcal{S}}(\bm{x})$; that is,  $Q_{\mcal{S}}(\bm{x}_{\partial\mcal{U}}) = N^{-1}\sum_{\mu=1}^N \prod_{i\in\partial\mcal{U}}\delta (x_i,\mathrm{s}_i^{(\mu)})$: 
\begin{align}
m_{\mathcal{T}}(\mcal{U};\mcal{S}) := \frac{1}{N}\sum_{\mu=1}^N \sum_{\bm{x}_{\mcal{U}}}f(\bm{x}_{\mcal{T}})
P_{\theta}(\bm{x}_{\mcal{U}}\mid\mbf{s}_{\partial\mcal{U}}^{(\mu)}).
\label{eq:smci}
\end{align}
Equation (\ref{eq:smci}) is the form of the sample average of the conditional expectation of $f(\bm{x}_{\mcal{T}})$, 
\begin{align}
f_{\mcal{T}, \mcal{U}}(\mbf{s}_{\partial\mcal{U}}^{(\mu)}) 
:= \sum_{\bm{x}_{\mcal{U}}}f(\bm{x}_{\mcal{T}})P_{\theta}(\bm{x}_{\mcal{U}}\mid\mbf{s}_{\partial\mcal{U}}^{(\mu)}).
\label{eq:conditional_expectatoin_smci}
\end{align}
Such transformation is known as Rao-Blackwellization~\cite{liu2001}. 
The simplest SMCI is the first-order SMCI (1-SMCI) method wherein the sum region is identical to the target region. 
For example, consider the 1-SMCI method to approximate the expectation of $f(x_i) = x_i$ in the case of $\mcal{X} = \{-1,+1\}$.
Because the target and sum regions are $\mcal{T} = \mcal{U} = \{i\}$, 
and the sample region $\partial \mcal{U} = \partial i =\{j\mid(i,j)\in\mcal{E}, j\in \mcal{V} \setminus \{i\}\}$, 
the conditional distribution is written as
\begin{align*}
P_{\theta}(\bm{x}_{\mcal{U}}\mid\mbf{s}_{\partial\mcal{U}}^{(\mu)}) = P_{\theta}(x_i \mid\mbf{s}_{\partial {i}}^{(\mu)})
= \frac{ \exp\big[\big( h_i + \sum_{j \in \partial i} J_{i,j} \mrm{s}_j^{(\mu)}\big)x_i \big]}
{2 \cosh \big( h_i + \sum_{j \in \partial i} J_{i,j} \mrm{s}_j^{(\mu)}\big)}.
\end{align*}
Therefore, equation (\ref{eq:smci}) becomes 
\begin{align*}
m_i(i, \mcal{S}) = \frac{1}{N}\sum_{\mu=1}^N \sum_{x_i \in \{-1,+1\} }x_i P_{\theta}(x_i \mid\mbf{s}_{\partial i}^{(\mu)})
= \frac{1}{N}\sum_{\mu=1}^N \tanh\Big( h_i + \sum_{j \in \partial i} J_{i,j} \mrm{s}_j^{(\mu)}\Big).
\end{align*} 

The estimator of equation (\ref{eq:smci}) is unbiased 
because $\sexp{m_{\mathcal{T}}(\mathcal{U};\mathcal{S})} = \mexp{f(\bm{x}_{\mcal{T}})}$, where 
\begin{align}
\sexp{A} := 
\sum_{\mbf{s}^{(1)}} P_{\theta}(\mbf{s}^{(1)} ) \sum_{\mbf{s}^{(2)}} P_{\theta}(\mbf{s}^{(2)})\cdots \sum_{\mbf{s}^{(N)}}  P_{\theta}(\mbf{s}^{(N)}) A
\label{eq:sampleset_average}
\end{align}
denotes the average over the sample set $\mcal{S}$. 
From this unbiasedness and the central limit theorem, 
the distribution of the SMCI estimator is found to asymptotically converge to a normal distribution whose mean is the true expectation for a sufficient large $N$. 
The following two asymptotic properties have been proved~\cite{yasuda2015,yasuda2021}: 
for a given sample set, (i) SMCI is statistically more accurate than the standard MCI, and 
(ii) the accuracy of $m_\mathcal{T}(\mathcal{U}_b;\mathcal{S})$ is statistically higher than that of $m_\mathcal{T}(\mathcal{U}_a;\mathcal{S})$ when $\mcal{T}\subseteq\mcal{U}_a\subseteq\mcal{U}_b$. 
From property (ii), the expansion of the sum region is theoretically guaranteed to statistically improve the approximation accuracy.
However, in general, the expansion causes the exponential increase of the computational cost, except in some special cases (e.g., when the graph is a tree). 
Assuming the cost of the evaluation of $f(\bm{x}_{\mcal{T}})$ is $O(F)$ and $\mcal{X}$ is a discrete space, 
the cost of the evaluation of $m_{\mcal{T}}(\mcal{U};\mcal{S})$ is estimated as $O(N F|\mcal{X}|^u)$, where $u = |\mcal{U}|$; 
here, the cost of $O(|\mcal{X}|^u)$ is from the multiple summation over $\bm{x}_{\mcal{U}}$. 
Note that the cost of sampling is not considered in this cost evaluation. 

\begin{figure}[t]
\centering
\includegraphics[width=0.35\linewidth]{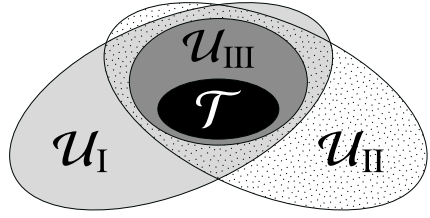}
\caption{Illustration of the inclusion relation of the sum regions $\mcal{U}_{\mrm{I}}$, $\mcal{U}_{\mrm{II}}$, and $\mcal{U}_{\mrm{III}}$. }
\label{fig:3-regions}
\end{figure}

Suppose that we have three different sum regions, $\mcal{U}_{\mrm{I}}$, $\mcal{U}_{\mrm{II}}$, and $\mcal{U}_{\mrm{III}}$,
for a fixed target region $\mcal{T}$, and that their inclusion relation is as illustrated in Figure \ref{fig:3-regions}. 
From property (ii), $m_\mathcal{T}(\mathcal{U}_{\mrm{I}};\mathcal{S})$ and $m_\mathcal{T}(\mathcal{U}_{\mrm{II}};\mathcal{S})$ are statistically more accurate than $m_\mathcal{T}(\mathcal{U}_{\mrm{III}};\mathcal{S})$ because $\mcal{U}_{\mrm{III}} \subseteq \mcal{U}_{\mrm{I}}$ and $\mcal{U}_{\mrm{III}} \subseteq \mcal{U}_{\mrm{II}}$. 
However, the accuracies of $m_\mathcal{T}(\mathcal{U}_{\mrm{I}};\mathcal{S})$ and $m_\mathcal{T}(\mathcal{U}_{\mrm{II}};\mathcal{S})$ 
is not theoretically compared because the relation of the two sum regions is neither $\mcal{U}_{\mrm{I}} \subseteq \mcal{U}_{\mrm{II}}$ 
nor $\mcal{U}_{\mrm{I}} \supseteq \mcal{U}_{\mrm{II}}$. 
In this case, it is difficult to select the best one.
To address this problem, we consider an alternative approach in the subsequent section, 
in which a new estimator is created by combining multiple SMCI estimators obtained from different sum regions.  

\section{Composite Spatial Monte Carlo Integration} \label{sec:CSMCI} 

In this section, to improve the approximation accuracy, we propose a new estimator based on the combination of multiple SMCI estimators obtained from different sum regions. 
Subsequently, we investigate the validity of the proposed method from theoretical and numerical perspectives 
in sections \ref{ssec:theoretical} and \ref{ssec:numerical}, respectively.

Suppose that we have $K$ different sum regions that cover the target region $\mathcal{T}$, $\mathcal{U}_1,\mathcal{U}_2,\cdots,\mathcal{U}_K\supseteq\mathcal{T}$, 
to evaluate the true expectation, $m_{\mcal{T}}^{\mrm{true}}:=\mrm{E}_{\theta}[f(\bm{x}_{\mcal{T}})] $, in equation (\ref{eq:expectation_f(x_T)}) based on SMCI. 
For the sample set $\mathcal{S}$, 
according to equation (\ref{eq:smci}), we can obtain $K$ different SMCI estimators for $\mrm{E}_{\theta}[f(\bm{x}_{\mcal{T}})] $ using the $K$ different sum regions as 
\begin{align}
m_{\mcal{T}}^{(k)} := m_{\mcal{T}}(\mcal{U}_k; \mcal{S}) = \frac{1}{N}\sum_{\mu=1}^Nf_{\mcal{T}, \mcal{U}_k} (\mbf{s}_{\partial\mcal{U}_k}^{(\mu)}), 
\quad k = 1,2,\ldots, K,	
\end{align}
where $f_{\mcal{T}, \mcal{U}_k} (\mbf{s}_{\partial\mcal{U}_k}^{(\mu)})$ is the conditional expectation for the $k$th sum region 
defined in equation (\ref{eq:conditional_expectatoin_smci}). 
We collectively denote the $K$ different SMCI estimators by the vector
\begin{align}
\bm{m}_{\mcal{T}}:=\big(m_{\mcal{T}}^{(1)},m_{\mcal{T}}^{(2)},\ldots,m_{\mcal{T}}^{(K)}\big)^{\mrm{t}} 
\in \mathbb{R}^K.  
\label{eq:smci_estimator_vector}
\end{align}
This vector is the random vector whose elements (i.e., $K$ different SMCI estimators) are random variables. 

As mentioned in section \ref{sec:SMCI}, the SMCI estimators are unbiased; therefore, the mean vector of $\bm{m}_{\mcal{T}}$ is $\bm{\mu} := m_{\mcal{T}}^{\mrm{true}} \bm{1}_K \in \mathbb{R}^K$, where $\bm{1}_K$ is the $K$-dimensional one vector. 
The covariance matrix of $\bm{m}_{\mcal{T}}$, $\bm{\Sigma} \in\mathbb{R}^{K\times K}$, is defined as $\Sigma_{k,k'} := \sexp{m_{\mcal{T}}^{(k)}m_{\mcal{T}}^{(k')}}
- \sexp{m_{\mcal{T}}^{(k)}} \sexp{m_{\mcal{T}}^{(k')}}$, where $\mrm{E}_{\mcal{S}}[\cdots]$ denotes the average over the sample set defined in equation (\ref{eq:sampleset_average}). 
It should be noted that the $K$ different SMCI estimators are not statistical independents of each other because they share the same sample set.
Here, $\bm{\Sigma}$ has been assumed to be positive definite. 
Since $\sexp{f_{\mcal{T}, \mcal{U}_k}(\mbf{s}_{\partial\mcal{U}_k}^{(\mu)})} 
=\mexp{f_{\mcal{T}, \mcal{U}_k}(\bm{x}_{\partial\mcal{U}_k})}= m_{\mcal{T}}^{\mrm{true}}$ for all $k$ and $\mu$, 
\begin{align}
\Sigma_{k,k'} = \frac{1}{N} \MEXP{ 
\big(f_{\mcal{T}, \mcal{U}_k}(\bm{x}_{\partial\mcal{U}_k}) - m_{\mcal{T}}^{\mrm{true}}\big)
\big(f_{\mcal{T}, \mcal{U}_{k'}}(\bm{x}_{\partial\mcal{U}_{k'}}) - m_{\mcal{T}}^{\mrm{true}}\big)
}
\label{eq:def_cov_matrix}
\end{align}
is obtained, where $\Sigma_{k,k'}$ vanishes at a speed of $O(N^{-1})$.
Using an error vector $\bm{\varepsilon} \in \mathbb{R}^K$, $\bm{m}_{\mcal{T}}$ is represented as 
\begin{align}
\bm{m}_{\mcal{T}} = m_{\mcal{T}}^{\mrm{true}} \bm{1}_K + \bm{\varepsilon},
\label{eq:reg_form_base}
\end{align}
The error vector is the random vector whose the mean vector is the $K$-dimensional zero vector, $\bm{0}_K$, and whose the covariance matrix is $\bm{\Sigma}$. 
When $N$ is sufficiently large, per the (multivariate) central limit theorem, 
the distribution of the error vector asymptotically converges to a normal distribution $\mcal{N}(\bm{\varepsilon} \mid \bm{0}_K, \bm{\Sigma})$; 
while, when $N$ is finite, the distribution is generally not a normal distribution.

Regarding the relation of equation (\ref{eq:reg_form_base}) as a linear regression problem, 
GLS provides the framework to create a new estimator for $m_{\mcal{T}}^{\mrm{true}}$, using the $K$ different SMCI estimators.
In the context of GLS~\cite{aitken1936,greene2003,lee2018}, the ``unknown'' true expectation in equation (\ref{eq:reg_form_base}) is regarded as regression coefficient 
$\alpha$; that is,  
\begin{align}
\bm{m}_{\mcal{T}} = \alpha\bm{1}_K + \bm{\varepsilon}.
\label{eq:reg_form}
\end{align}
Subsequently, as the elements of the error vector are correlated, 
the optimal regression coefficient is obtained in terms of the minimization of the Mahalanobis distance, i.e., 
\begin{align}
\hat{\alpha}_{\mcal{T}} := \argmin_{\alpha} \big( \bm{m}_{\mcal{T}} - \alpha \bm{1}_K\big)^{\mrm{t}} \bm{\Sigma}^{-1}
\big( \bm{m}_{\mcal{T}} - \alpha\bm{1}_K\big).
\label{eq:Mahalanobis_distance}
\end{align}
The resulting estimation $\hat{\alpha}_{\mcal{T}}$ is regarded as the approximation of $m_{\mcal{T}}^{\mrm{true}}$. 
The minimization of equation (\ref{eq:Mahalanobis_distance}) yields 
\begin{align}
\hat{\alpha}_{\mcal{T}} = \bm{c}^{\mrm{t}}\bm{m}_{\mcal{T}},
\label{eq:GLS_estimator}
\end{align}
where  $\bm{c} \in \mathbb{R}^K$ is the coefficient defined by 
\begin{align}
\bm{c} := \frac{1}{\Omega(\bm{\Sigma}^{-1})}\bm{\Sigma}^{-1}\bm{1}_K. 
\label{eq:coefficient_c}
\end{align} 
Here, for $\bm{A} \in \mathbb{R}^{K \times K}$, $\Omega(\bm{A}) := \bm{1}_K^{\mrm{t}} \bm{A} \bm{1}_K \in \mathbb{R}$ 
denotes the sum of all elements of the assigned matrix. 
$\hat{\alpha}_{\mcal{T}}$ is the proposed estimator in this paper (hereinafter referred to as the CSMCI estimator). 
The CSMCI estimator is the linear combination of the $K$ SMCI estimators. 
The CSMCI estimator is unbiased because $\sexp{\hat{\alpha}_{\mcal{T}}} = m_{\mcal{T}}^{\mrm{true}}$,  
and its variance is 
\begin{align}
\mrm{V}_{\mcal{S}}[\hat{\alpha}_{\mcal{T}}] = \bm{c}^{\mrm{t}} \bm{\Sigma}\bm{c}=
\Omega \big(\bm{\Sigma}^{-1} \big)^{-1} = O(N^{-1}),
\label{eq:variance_GLS_estimator}
\end{align}
where $\mrm{V}_{\mcal{S}}[A]:=\sexp{A^2} - \sexp{A}^2$. 
The CSMCI estimator given in equation (\ref{eq:GLS_estimator}) can be obtained from an alternative approach based on Lagrange multipliers;
the detailed of explanation for this is provided in Appendix\ref{app:lagrange}.

The CSMCI estimator requires the evaluation of the covariance matrix $\bm{\Sigma}$.
However, the covariance matrix includes intractable expectations; 
therefore, practically it has to be approximated (the so-called feasible GLS~\cite{greene2003}).  
In this paper, it is simply approximated by the (unbiased) sample covariance matrix; that is, 
\begin{align}
\bm{\Sigma} \approx \bm{\Sigma}_{\mrm{app}}
:=\frac{1}{N}\Big(\frac{1}{N-1}\sum_{\mu=1}^N \bm{r}_{\mu}\bm{r}_{\mu}^{\mrm{t}} \Big),
\label{eq:cov-mat_approximation}
\end{align}
where $\bm{r}_{\mu} \in \mathbb{R}^K$ is the vector whose $k$th element is defined as 
\begin{align*}
r_{\mu, k} := f_{\mcal{T}, \mcal{U}_k} (\mbf{s}_{\partial\mcal{U}_k}^{(\mu)}) - m_{\mcal{T}}^{(k)}.
\end{align*}
The approximation error in the parentheses of $\bm{\Sigma}_{\mrm{app}}$ in equation (\ref{eq:cov-mat_approximation}) decreases at a speed proportional to $O(N^{-1/2})$; thus, the approximation error of $\bm{\Sigma}_{\mrm{app}}$ decreases at a speed proportional to $O(N^{-3/2})$, i.e., $\bm{\Sigma} = \bm{\Sigma}_{\mrm{app}} +O(N^{-3/2})$.
Using $\bm{\Sigma}_{\mrm{app}}$, the CSMCI estimator $\hat{\alpha}_{\mcal{T}}$ is approximated by 
\begin{align}
\hat{\alpha}_{\mcal{T}}\approx \bm{c}_{\mrm{app}}^{\mrm{t}}\bm{m}_{\mcal{T}},
\label{eq:CSMCI_estimator}
\end{align}
where
\begin{align*}
\bm{c}_{\mrm{app}}:=\frac{1}{\Omega(\bm{\Sigma}_{\mrm{app}}^{-1})}\bm{\Sigma}_{\mrm{app}}^{-1}\bm{1}_K.
\end{align*}
The estimator of equation (\ref{eq:CSMCI_estimator}) is referred to as quasi-CSMCI (qCSMCI) estimator.
In practice, we use the qCSMCI estimator instead of the CSMCI estimator of equation (\ref{eq:GLS_estimator}). 
Since $\bm{\Sigma} = \bm{\Sigma}_{\mrm{app}} +O(N^{-3/2})$, the qCSMCI estimator converges to the CSMCI estimator as $N$ increases. 
This implies that the qCSMCI estimator converses to $m_{\mcal{T}}^{\mrm{true}}$ in the limit of $N \to \infty$.

\subsection{Theoretical Validation} 
\label{ssec:theoretical} 

In this section, we discuss the validity of the CSMCI estimator in equation (\ref{eq:GLS_estimator}) without the covariance-matrix approximation in equation (\ref{eq:cov-mat_approximation}), from a theoretical perspective.  
From the result obtained in the Appendix\ref{app:lagrange}, 
it is found that the CSMCI estimator $\hat{\alpha}_{\mcal{T}}$ is the best estimator, from the perspective of variance, among all unbiased estimators obtained by linear combinations of $\bm{m}_{\mcal{T}}$; thus, it is called best linear unbiased estimator (BLUE). 
The same result can be obtained from the Gauss-Markov theorem in a rigorous manner~\cite{aitken1936,greene2003}. 

The fact of that the CSMSI estimator is the BLUE immediately leads to the following two important properties. 
The first is that the variance of the CMSCI estimator is a lower bound of the variances of the $K$ SMCI estimators, 
\begin{align}
\mrm{V}_{\mcal{S}}[\hat{\alpha}_{\mcal{T}}] \leq \min_{k=1,2,\ldots,K} \mrm{V}_{\mcal{S}}[m_{\mcal{T}}^{(k)}], 
\label{eq:collorary-1}
\end{align}
because the CSMCI estimator is the BLUE, and $m_{\mcal{T}}^{(k)}$ can be regarded as linear combination  
$m_{\mcal{T}}^{(k)} = \bm{v}^{\mrm{t}} \bm{m}_{\mcal{T}}$, the coefficients of which are $v_k = 1$, and the others are zero.
Equation (\ref{eq:collorary-1}) implies that the CMSCI estimator is guaranteed to improve the approximation accuracy.
The other one, which is described below, can be led from the same fact.
Consider $(K+1)$ SMCI estimators obtained by adding a new SMCI estimator, $m_{\mcal{T}}^{(K+1)}$, 
with a new sum region $\mcal{U}_{K+1} \supseteq \mcal{T}$ to $\bm{m}_{\mcal{T}}$: 
$\bm{m}_{\mcal{T}}^{+}:=\big(m_{\mcal{T}}^{(1)},m_{\mcal{T}}^{(2)},\ldots,m_{\mcal{T}}^{(K)}, m_{\mcal{T}}^{(K+1)}\big)^{\mrm{t}}$. 
We denote the CSMCI estimator for the $(K+1)$ SMCI estimators as $\hat{\alpha}_{\mcal{T}}^{+}$ obtained in a  similar manner to equation  (\ref{eq:GLS_estimator}).  The following inequality holds: 
\begin{align}
\mrm{V}_{\mcal{S}}[\hat{\alpha}_{\mcal{T}}^{+}] \leq \mrm{V}_{\mcal{S}}[\hat{\alpha}_{\mcal{T}}] .
\label{eq:collorary-2}
\end{align}
This inequality can be obtained from almost the same logic as that of equation (\ref{eq:collorary-1}) 
(i.e., $\hat{\alpha}_{\mcal{T}}$ can be regarded as a linear combination of $\bm{m}_{\mcal{T}}^{+}$ 
in which the coefficient for $m_{\mcal{T}}^{(K+1)}$ is set to zero).
Equation (\ref{eq:collorary-2}) implies that the accuracy of the CMSCI estimator is monotonically improved by adding a new SMCI estimator. 
The above two properties in equations (\ref{eq:collorary-1}) and (\ref{eq:collorary-2}) are not asymptotic properties; therefore, they are justified in a finite $N$.

Fortunately, a stronger claim is possible.
The CSMCI estimator is the BUE (or minimum variance unbiased (MVU) estimator),  
which means that the CSMCI estimator is the best among all possible unbiased estimators obtained from $\bm{m}_{\mcal{T}}$ (not necessarily linear).  
The explanation is as follows.
It is known that a GLS estimator is BUE when the distribution of the error vector $\bm{\varepsilon}$ is a normal distribution~\cite{lee2018}. 
As mentioned in the previous section, the distribution of the error vector asymptotically converges to the normal distribution for a sufficiently large $N$. 
Therefore, we can conclude that the CSMCI estimator is asymptotically the BUE for a sufficiently large $N$. 
This fact can be understood through the perspective of the maximum likelihood (ML) estimation~\cite{lee2018}, 
a brief explanation of which is described in Appendix\ref{app:BUE}. 
Most recently, a surprising theorem, the \textit{modern} Gauss-Markov theorem, was reported~\cite{hansen2022}. 
According to the \textit{modern} Gauss-Markov theorem, a GLS estimator is the BUE regardless of the distribution of the error vector. 
This means that the CSMCI estimator is the BUE in a finite $N$. 
From the above discussion, it can be concluded that the CSMCI estimator is always the BUE.

We have discussed the validity of the proposed CSMCI estimator in equation (\ref{eq:GLS_estimator}) from the theoretical perspective. 
However, it is not a practical estimator because it needs to treat the intractable covariance matrix.  
Therefore, in practice, we have to use the CSMCI estimator together with a covariance-matrix approximation, e.g., 
the qCSMCI estimator in equation (\ref{eq:CSMCI_estimator}).  
As mentioned in section \ref{sec:CSMCI}, the qCSMCI estimator converses to the CSMCI estimator in the limit of $N \to \infty$.
However, the above theoretical results are not guaranteed in the qCSMCI estimator in the case of a finite $N$. 
Thus, they are nothing more than \textit{expectations} for the qCSMCI estimator in that case.
Furthermore, the unbiasedness of the qCSMCI estimator is not theoretically guaranteed.
In the subsequent section, we demonstrate the validity of the qCSMCI estimator through numerical experiments.

\subsection{Experimental Validation} 
\label{ssec:numerical} 

\begin{figure*}[t]
\centering
\begin{tabular}{ccc}
\begin{minipage}[b]{0.3\linewidth}
\centering
\includegraphics[width=0.8\linewidth]{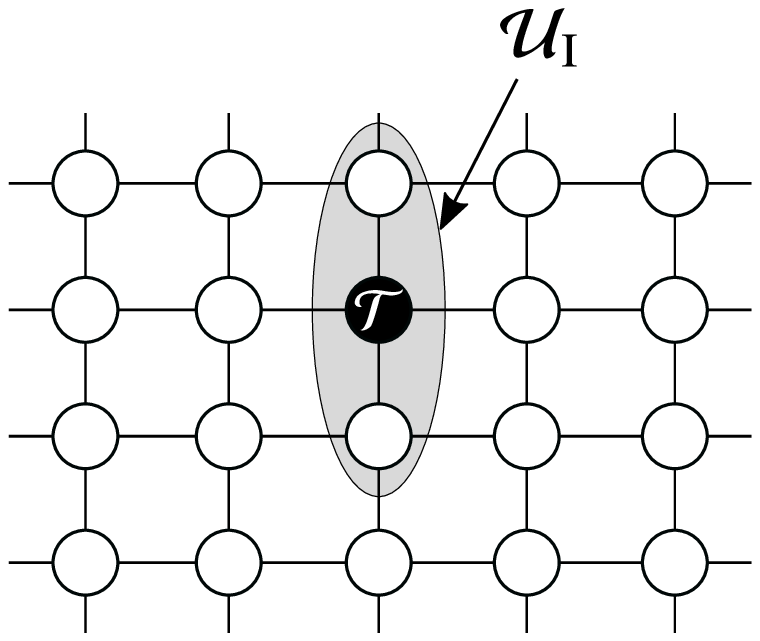} \\
(a)
\end{minipage}
\begin{minipage}[b]{0.3\linewidth}
\centering
\includegraphics[width=0.8\linewidth]{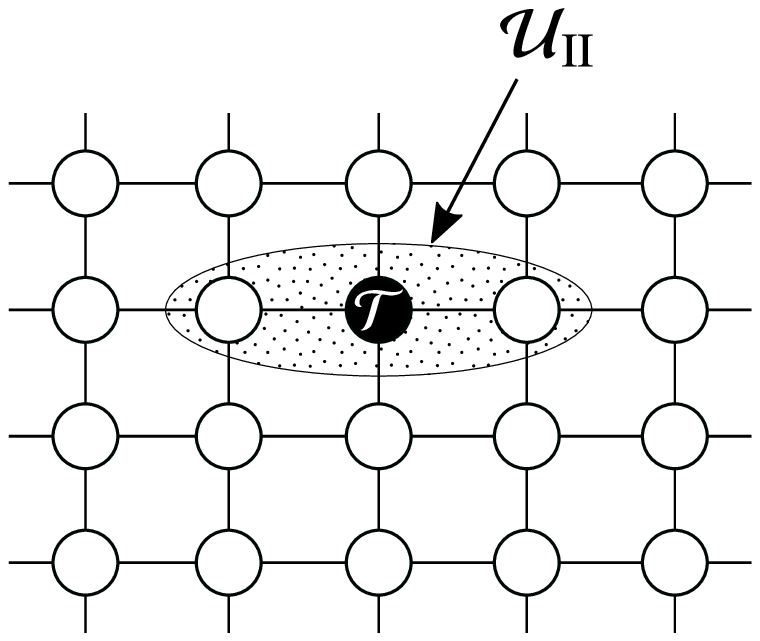} \\
(b)
\end{minipage}
\begin{minipage}[b]{0.3\linewidth}
\centering
\includegraphics[width=0.8\linewidth]{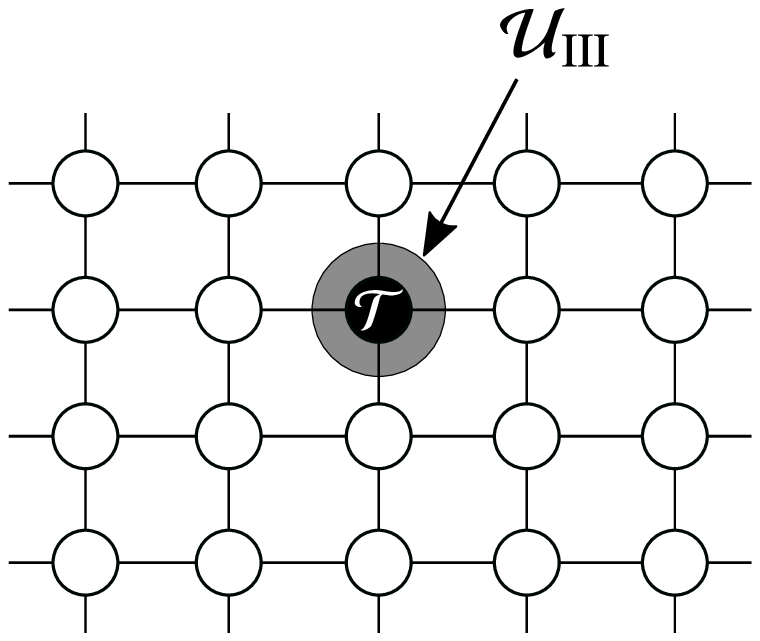} \\
(c)
\end{minipage}
\end{tabular}
\caption{Three different sum regions, namely, $\mcal{U}_{\mrm{I}}$, $\mcal{U}_{\mrm{II}}$, and $\mcal{U}_{\mrm{III}}$, for $\mcal{T}=\{i\}$.}
\label{fig:sum_region_i}
\end{figure*}

In this section, we demonstrate the validity of the qCSMCI estimator in equation (\ref{eq:CSMCI_estimator}) through numerical experiments. 

\subsubsection{Experiments on small Ising model}
\label{sssec:experiment_small}

We demonstrate the effectiveness of the qCSMCI estimator in a small Ising model defined on a $4\times 5$ square grid graph with the periodic boundary condition (i.e., a torus graph). 
In the small system, because various statistical quantities can be can numerically evaluated, we can precisely examine numerical properties of the qCSMCI estimator. 
In the Ising model, the sample space of the random variable was $\mathcal{X} = \{-1,+1\}$, and $h_i$ and $J_{i,j}$ were independently drawn from an uniform distribution in the interval $[-1/T, +1/T]$ (here, $T$ corresponds to the temperature).

On the Ising model, we evaluate uni-variable expectations; that is, $\mcal{T} = \{i\}$ and $f(x_i) = x_i$. 
For the target region, we considered three different sum regions: $\mcal{U}_{\mrm{I}}$, $\mcal{U}_{\mrm{II}}$, and $\mcal{U}_{\mrm{III}}$. 
Here, $\mcal{U}_{\mrm{III}}$ is the same as the target region, 
while $\mcal{U}_{\mrm{I}}$ and $\mcal{U}_{\mrm{II}}$ cover up to vertical and horizontal nearest-neighbor variables, respectively (see Figure \ref{fig:sum_region_i}). 
Since $\mcal{U}_{\mrm{I}}, \mcal{U}_{\mrm{II}} \supseteq \mcal{U}_{\mrm{III}} = \mcal{T}$, 
the SMCI estimators for $\mcal{U}_{\mrm{I}}$ and $\mcal{U}_{\mrm{II}}$ are guaranteed to be more accurate than that for $\mcal{U}_{\mrm{III}}$ 
from the theory of SMCI~\cite{yasuda2015,yasuda2021}. 
The SMCI estimators with the three sum regions, namely, $\mcal{U}_{\mrm{I}}$,  $\mcal{U}_{\mrm{II}}$, and $\mcal{U}_{\mrm{III}}$, 
are referred to as ``SMCI-I,'' ``SMCI-II,'' and ``SMCI-III,'' respectively. 
For the three SMCI estimators, we considered two different qCSMCI estimators: 
the first one using SMCI-I and SMCI-II (referred to as ``qCSMCI-I+II''), and the other one using all three SMCI estimators (referred to as  ``qCSMCI-all''). 
In the experiments, the sample set consisting of $N$ sample points was obtained using Gibbs sampling, in which the number of the Markov Chain Monte Carlo (MCMC) steps $r$, in both burn-in time and sampling interval, was fixed to $r = 50$.  
The approximation accuracy was measured by the mean absolute error (MAE) defined by
\begin{align}
\frac{1}{n} \sum_{i \in \mcal{V}} \big| \mexp{x_i} - \mrm{E}_{\mrm{app}}[x_i] \big|,
\label{eq:MAE_E[xi]}
\end{align} 
where $\mrm{E}_{\mrm{app}}[x_i]$ is a corresponding estimator obtained from the SMCI or qCSMCI method.

\begin{figure*}[t]
\centering
\begin{tabular}{cc}
\begin{minipage}[b]{0.45\linewidth}
\centering
\includegraphics[width=0.8\linewidth]{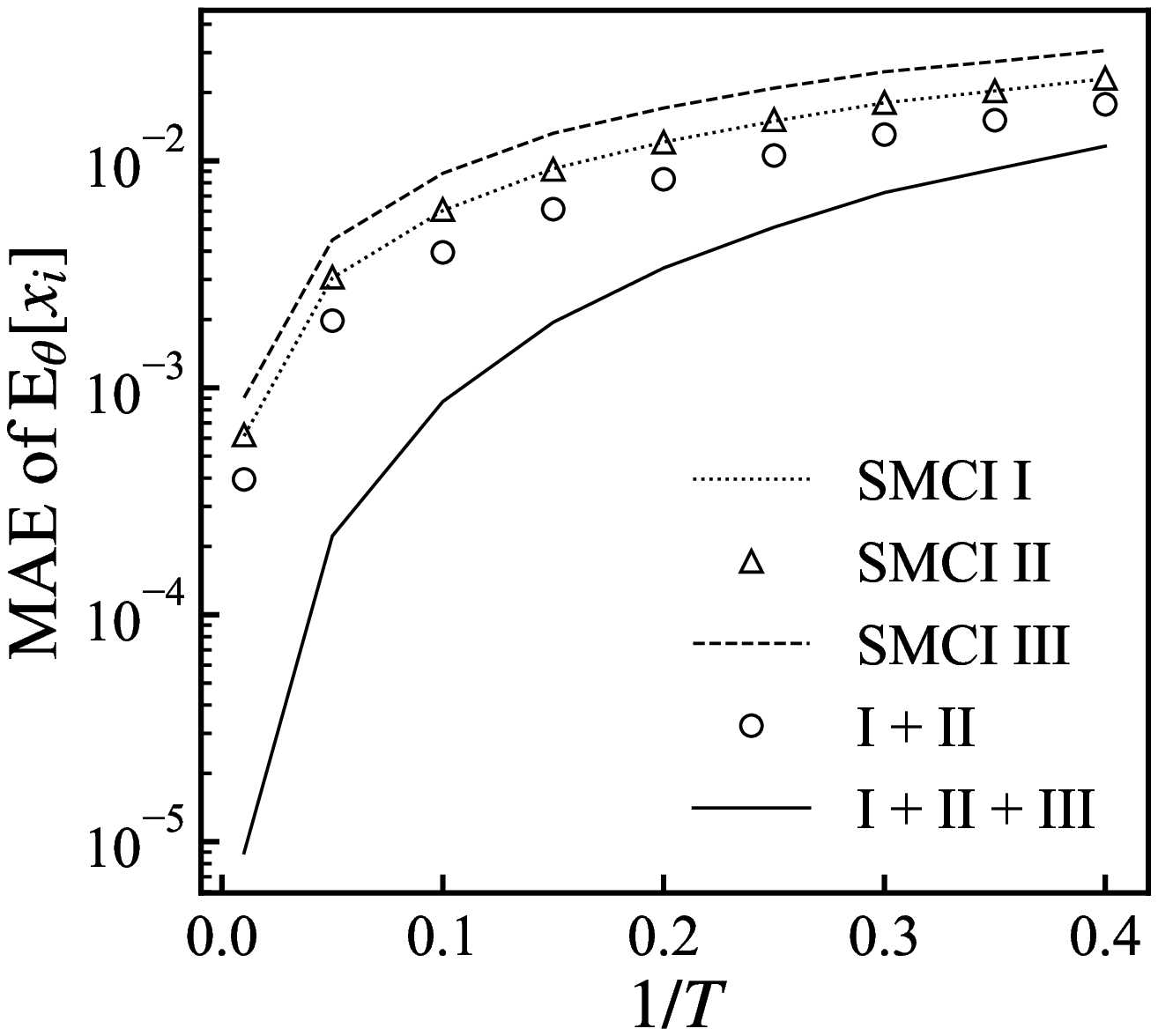} \\
\hspace*{0.89cm}(a)
\end{minipage}
\begin{minipage}[b]{0.45\linewidth}
\centering
\includegraphics[width=0.8\linewidth]{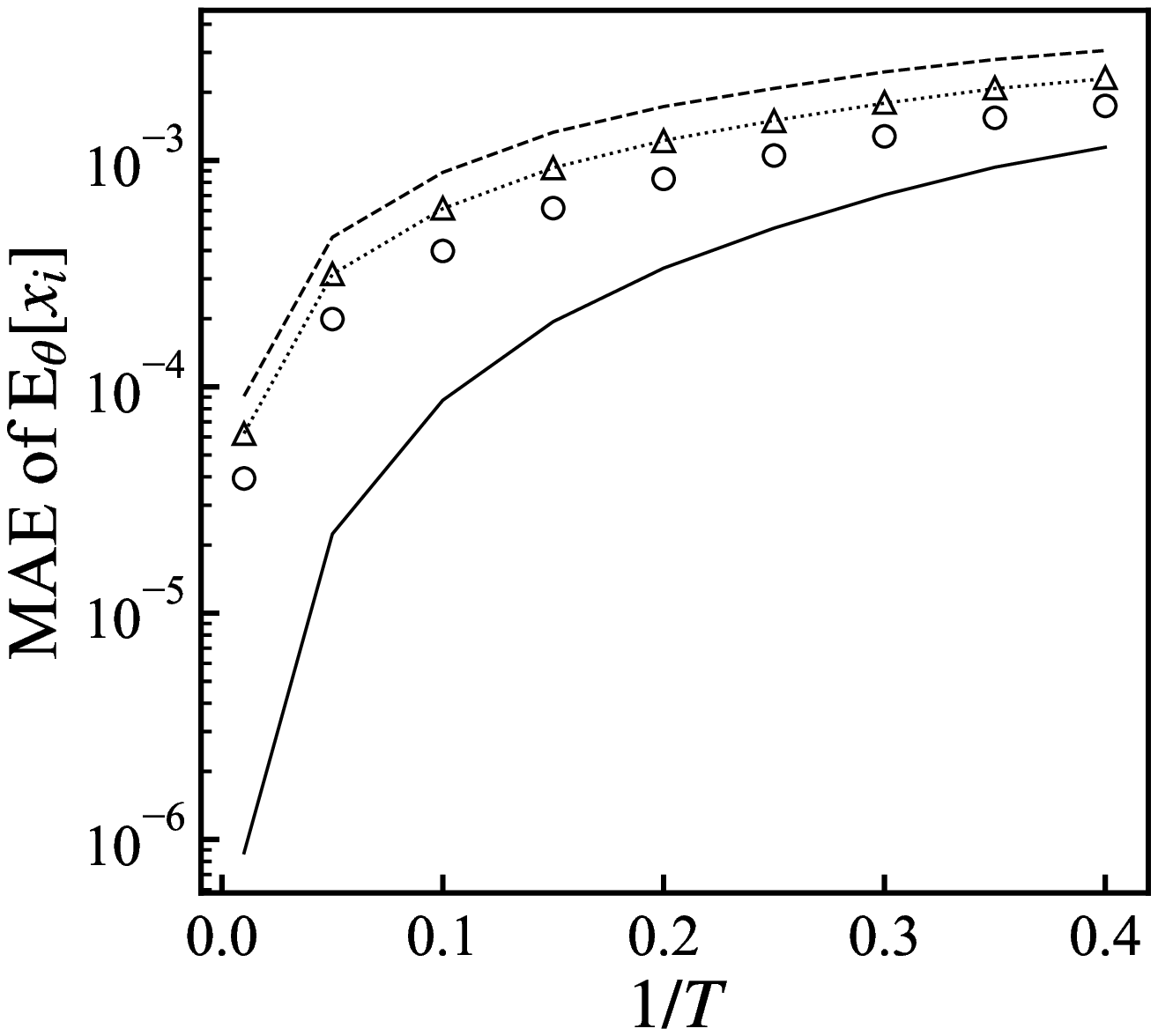} \\
\hspace*{0.89cm}(b)
\end{minipage}
\end{tabular}
\caption{MAEs of the evaluation of $\mexp{x_i}$ for $i \in \mcal{V}$ versus $1/T$ when (a) $N = 100$ and (b) $N = 10000$.
These plots are the average of $1000$ experiments.}
\label{fig:MAE_versus_1/T}
\end{figure*}

\begin{figure*}[t]
\centering
\begin{tabular}{cc}
\begin{minipage}[b]{0.45\linewidth}
\centering
\includegraphics[width=0.8\linewidth]{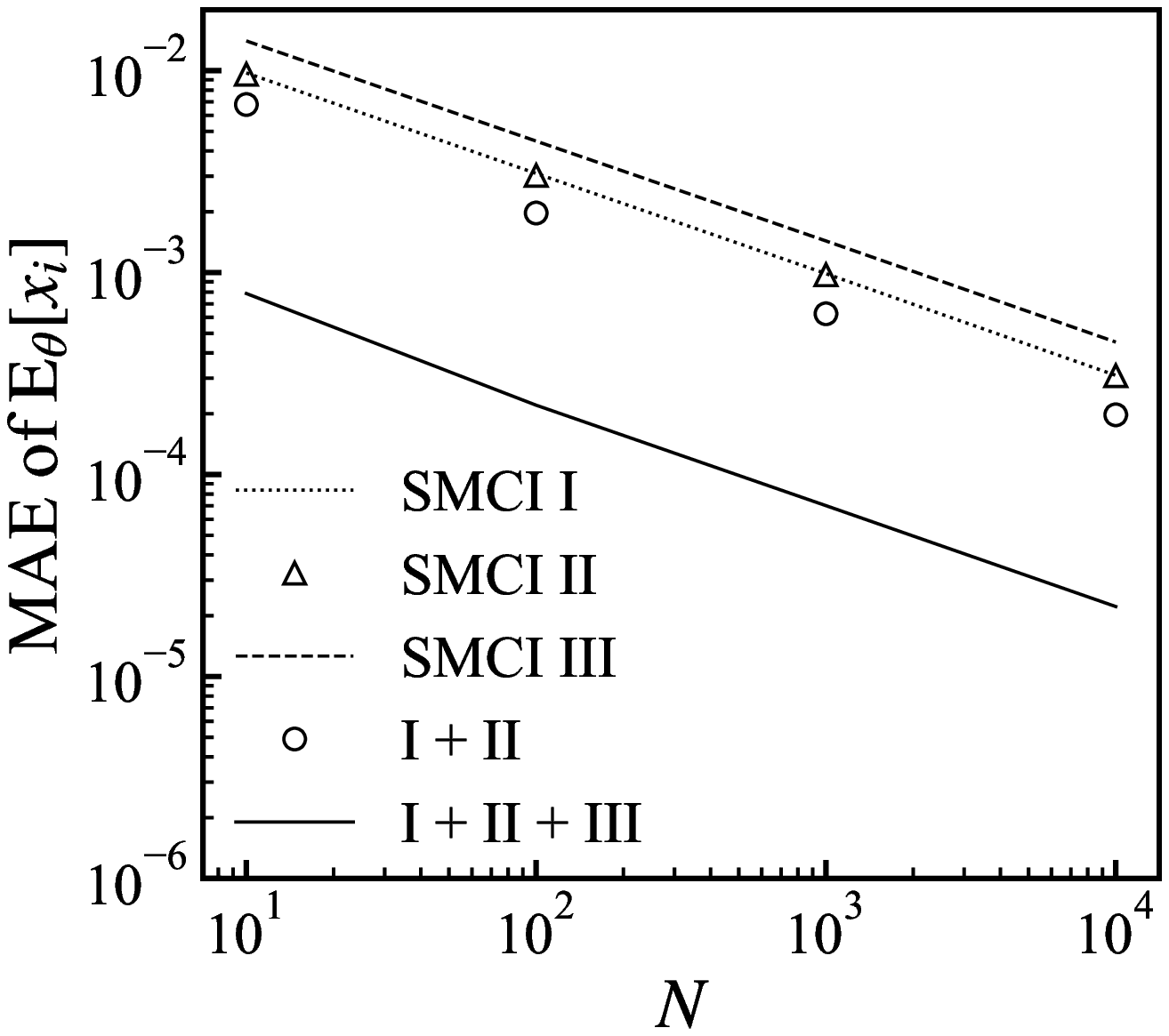} \\
\hspace*{0.89cm}(a)
\end{minipage}
\begin{minipage}[b]{0.45\linewidth}
\centering
\includegraphics[width=0.8\linewidth]{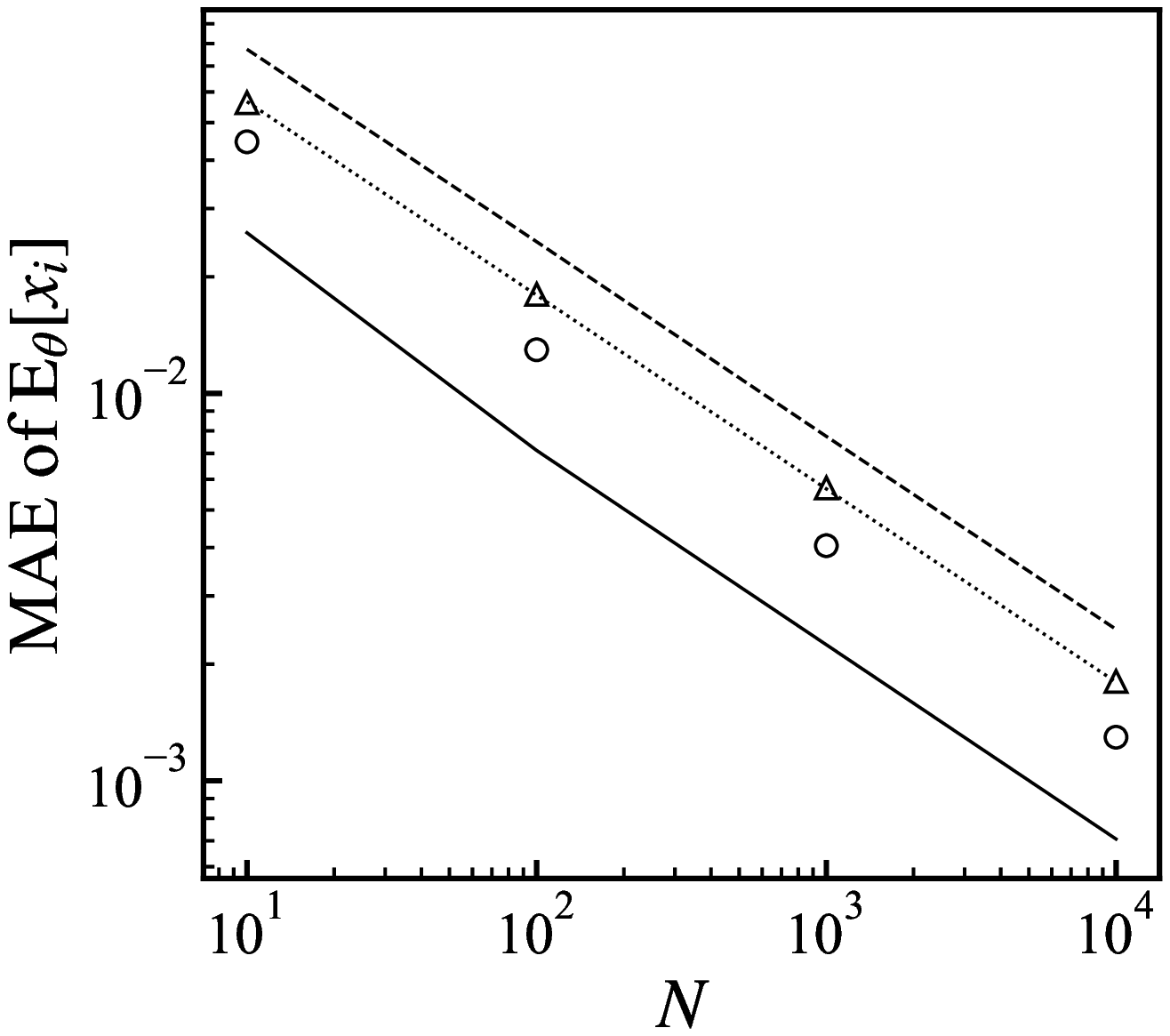} \\
\hspace*{0.89cm}(b)
\end{minipage}
\end{tabular}
\caption{MAEs of the evaluation of $\mexp{x_i}$ for $i \in \mcal{V}$ versus $N$ when (a) $1/T = 0.05$ and (b) $1/T = 0.3$.
These plots are the average of $1000$ experiments.}
\label{fig:MAE_versus_N}
\end{figure*}

\begin{figure*}[t]
\centering
\begin{tabular}{cc}
\begin{minipage}[b]{0.45\linewidth}
\centering
\includegraphics[width=0.8\linewidth]{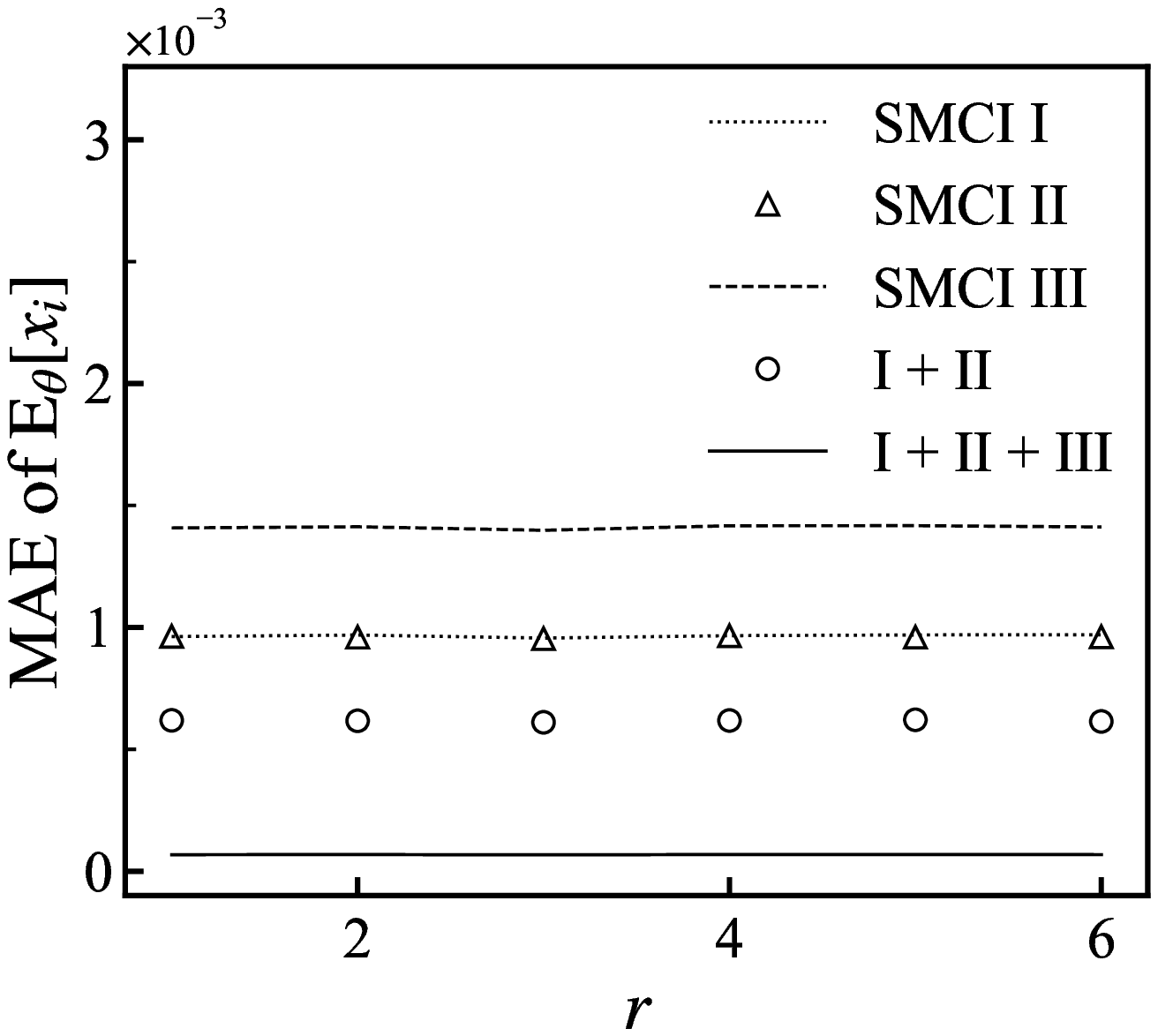} \\
\hspace*{0.6cm}(a)
\end{minipage}
\begin{minipage}[b]{0.45\linewidth}
\centering
\includegraphics[width=0.8\linewidth]{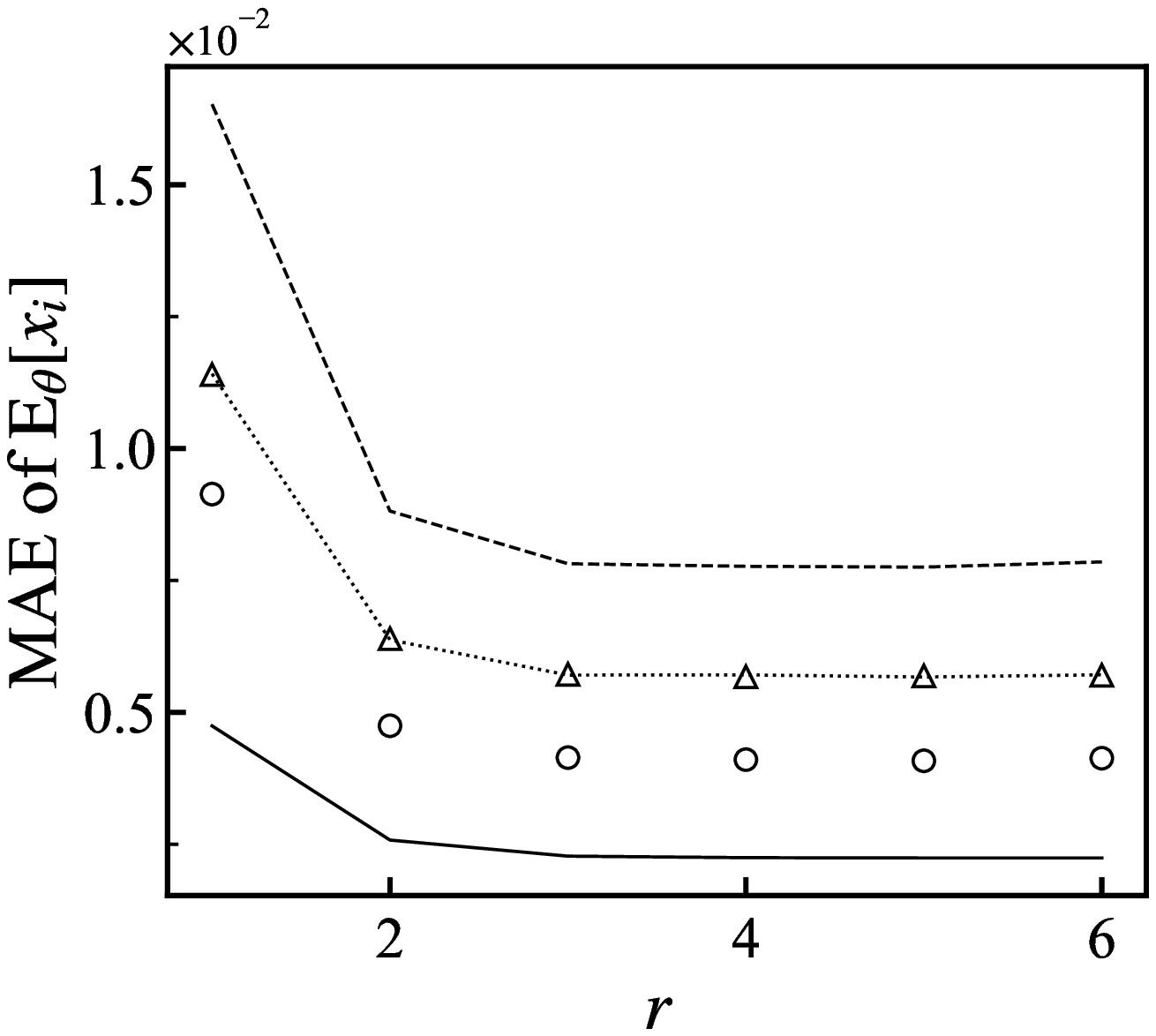} \\
\hspace*{0.8cm}(b)
\end{minipage}
\end{tabular}
\caption{MAEs of the evaluation of $\mexp{x_i}$ for $i \in \mcal{V}$ versus the number of the MCMC steps $r$ when (a) $1/T = 0.05$ and (b) $1/T = 0.3$.
These plots are the average of $1000$ experiments.}
\label{fig:MAE_versus_MCMCsteps}
\end{figure*}

Figure \ref{fig:MAE_versus_1/T} shows the plots of the MAEs against $1/T$.  
In the figure, ``I+II'' and ``I+II+III'' denote the results obtained from the qCSMCI-I+II and qCSMCI-all, respectively. 
Evidently, the numerical results agree with the \textit{expectations} from the theoretical results presented in section \ref{ssec:theoretical}; 
that is, the qCSMCI estimator improves the approximation accuracy 
and the accuracy of the qCMSCI estimator is monotonically improved by adding a new SMCI estimator.
Figure \ref{fig:MAE_versus_N} shows the plots of the MAEs against $N$.  
Clearly, the MAE of the qCSMCI estimators decrease at a speed approximately proportional to $O(N^{-1/2})$.
As mentioned in section \ref{ssec:theoretical}, it is not theoretically guaranteed that the qCSMCI estimator will be unbiased.
However, its behavior appears quite similar to that of an unbiased estimator.
In the experiments in Figures \ref{fig:MAE_versus_1/T} and \ref{fig:MAE_versus_N}, the number of the MCMC steps to sample the sample set was fixed to $r = 50$. 
In the next experiment, the dependency of the MAE on $r$ was investigated.
Figure \ref{fig:MAE_versus_MCMCsteps} shows the plots of the MAEs against $r$, in which $N=1000$ was fixed.
In Figure \ref{fig:MAE_versus_MCMCsteps}(b), the MAEs decrease as $r$ increases, and they saturate at few $r$ (around $r = 5$). 
The MAEs are relatively large in the small $r$ region, because the quality of the sampling is not good before the mixing time. 
In spite of the low quality of the sampling, in particular the qCSMCI-all, the qCMSCI estimators exhibit relatively small errors. 
This implies that the proposed method has the potential to correct the performance degradation caused by the low quality sampling.

\begin{figure*}[t]
\centering
\begin{tabular}{cc}
\begin{minipage}[b]{0.45\linewidth}
\centering
\includegraphics[width=0.8\linewidth]{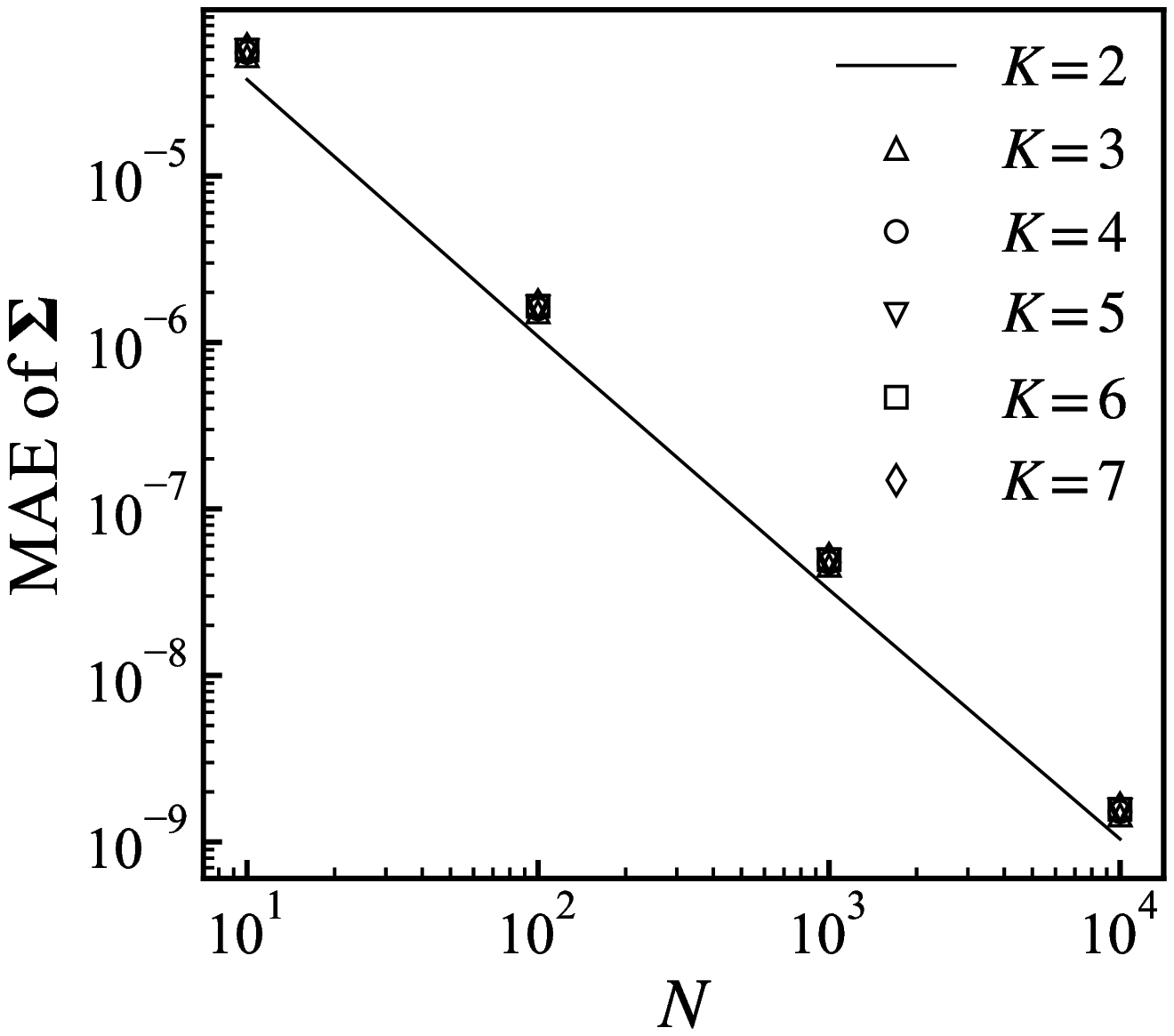} \\
\hspace*{0.86cm}(a)
\end{minipage}
\begin{minipage}[b]{0.45\linewidth}
\centering
\includegraphics[width=0.8\linewidth]{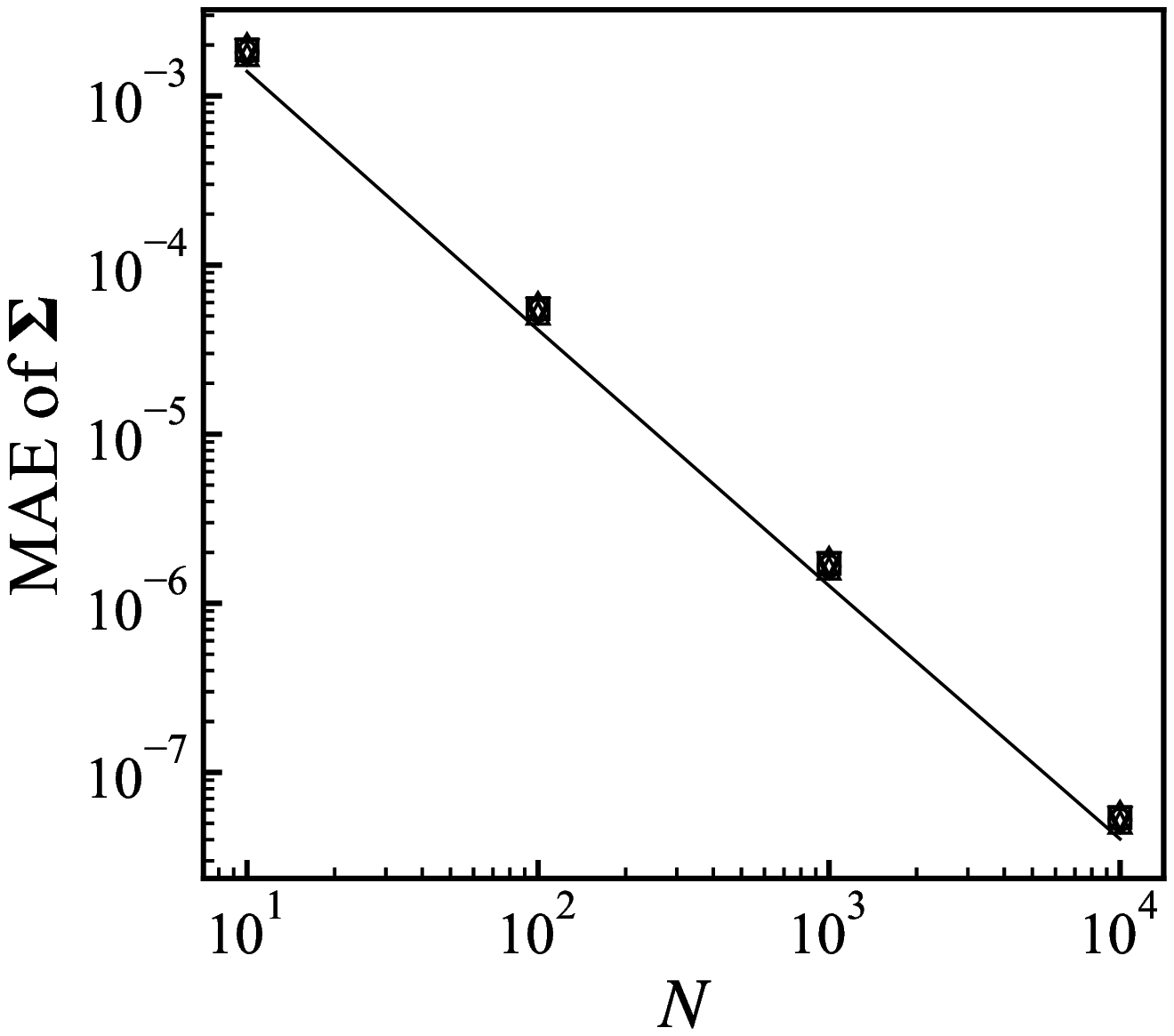} \\
\hspace*{0.9cm}(b)
\end{minipage}
\end{tabular}
\caption{MAEs of the approximation of $\bm{\Sigma}$ versus $N$ for various $K$ values when (a) $1/T = 0.05$ and (b) $1/T = 0.3$.
These plots are the average of $100$ experiments. The results for $K = 3,\ldots,7$ almost overlap.}
\label{fig:MAE_CovMat_versus_K}
\end{figure*}

The qCMSCI estimator is the approximation of the CSMCI estimator obtained from the approximation of $\bm{\Sigma} \approx \bm{\Sigma}_{\mrm{app}} \in \mathbb{R}^{K \times K}$.
To check the validity of the approximation, we investigate the dependency of the approximation error on $N$ and $K$.
For the target region $\mcal{T} = \{i\}$, in addition to $\mcal{U}_{\mrm{I}}$, $\mcal{U}_{\mrm{II}}$ and $\mcal{U}_{\mrm{III}}$, 
we introduce four different sum regions: $\mcal{U}_{\mrm{IV}}$, $\mcal{U}_{\mrm{V}}$, $\mcal{U}_{\mrm{VI}}$ and $\mcal{U}_{\mrm{VII}}$; 
the four sum regions cover the top, bottom, left and right nearest-neighbor variables, respectively (i.e., $\mcal{U}_{\mrm{IV}}\cup\mcal{U}_{\mrm{V}}=\mcal{U}_{\mrm{I}}$ and $\mcal{U}_{\mrm{VI}}\cup\mcal{U}_{\mrm{VII}}=\mcal{U}_{\mrm{II}}$).
To construct the CSMCI or qCSMCI estimator, we composite the SMCI estimators in the ascending order of the index of the sum region; for instance, the four SMCI estimators with $\mcal{U}_{\mrm{I}}$, $\mcal{U}_{\mrm{II}}$, $\mcal{U}_{\mrm{III}}$ and $\mcal{U}_{\mrm{IV}}$ are composited when $K=4$.
The accuracy of the approximation of the covariance matrix is measured by the MAE defined by
\begin{align}
\frac{1}{n}\sum_{i\in\mcal{V}}\frac{1}{K^2} \bigl\|\bm{\Sigma}^{(i)} - \bm{\Sigma}_{\mrm{app}}^{(i)}\bigr\|_{1},
\end{align}
where $\|\cdots\|_{1}$ denotes the element-wise $L_1$ matrix norm, 
and $\bm{\Sigma}^{(i)}$ and $\bm{\Sigma}_{\mrm{app}}^{(i)}$ are the covariance matrices obtained to evaluate $\mrm{E}_{\theta}[x_i]$ by the CSMCI and qCSMCI estimators, respectively.
Figure \ref{fig:MAE_CovMat_versus_K} shows the plots of the MAEs against $N$ for various $K$ values.
Although the MAEs slightly increase as $K$ increases, they are negligibly small, which implies an increase in $K$ almost does not change the approximation accuracy.
Also, the MAEs clearly decreased at a speed approximately proportional to $O(N^{-3/2})$ as mentioned in section \ref{ssec:theoretical}.

\subsubsection{Computational efficiency}

We discuss the computational efficiency of the qCSMCI estimator. 
From its definition, it is evident that the computational cost of the qCSMCI estimator is higher than that of each SMCI component. 
Suppose that the sum regions include at most a few variables.
In this case, empirically, the sampling cost is dominant in the total procedure to evaluate an SMCI estimator (it occupies more than 99\% in total in a case~\cite{sekimoto2021}) 
because the sampling procedure uses a costly pseudo-random-number generator. 
The difference of the evaluation costs of the qCSMCI estimator and its SMCI components can be small compared to the sampling cost.
Meanwhile, in the results presented in Figure \ref{fig:MAE_versus_N}, 
the SMCI-I or SMCI-II needs about 10--1000 times larger $N$ to achieve the same accuracy level of the qCSMCI-all. 
These facts support the computational efficiency of the qCSMCI estimator.
However, a rigorous comparison of the costs is not straightforward because the costs complicatedly depend on several aspects, such as the structure of model, choice of sampling method, and setting of sum regions.

\subsubsection{Experiments on large Ising model}
\label{sssec:experiment_large}

\begin{figure*}[t]
\centering
\begin{tabular}{cc}
\begin{minipage}[b]{0.3\linewidth}
\centering
\includegraphics[width=0.8\linewidth]{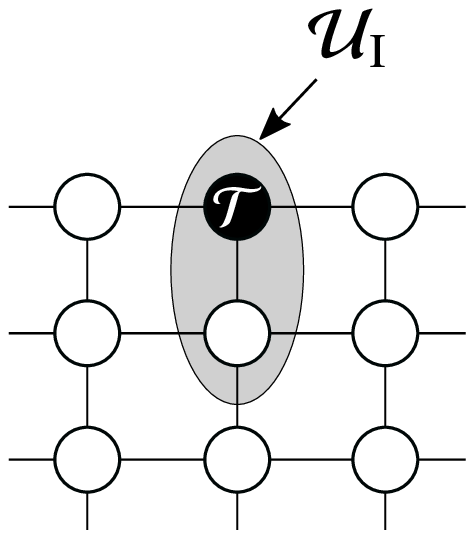} \\
(a)
\end{minipage}
\begin{minipage}[b]{0.3\linewidth}
\centering
\includegraphics[width=0.8\linewidth]{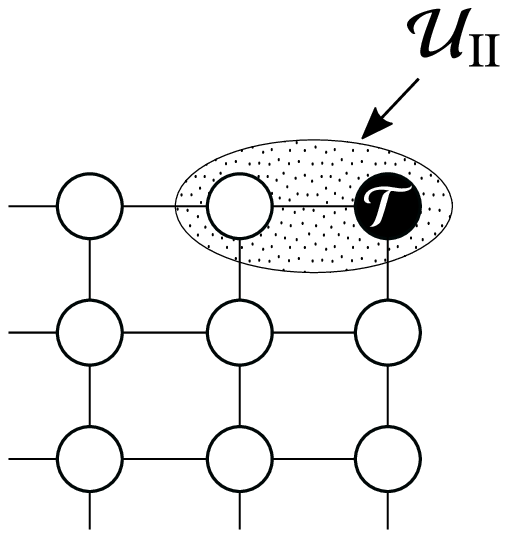} \\
(b)
\end{minipage}
\end{tabular}
\caption{Examples of sum regions, $\mcal{U}_{\mrm{I}}$ and $\mcal{U}_{\mrm{II}}$, at the edge cases. The overhang of the region is cut off (cf. Figure \ref{fig:sum_region_i}).}
\label{fig:sum_region_i_edge_case}
\end{figure*}
\begin{figure*}[t]
\centering
\begin{tabular}{cc}
\begin{minipage}[b]{0.45\linewidth}
\centering
\includegraphics[width=0.8\linewidth]{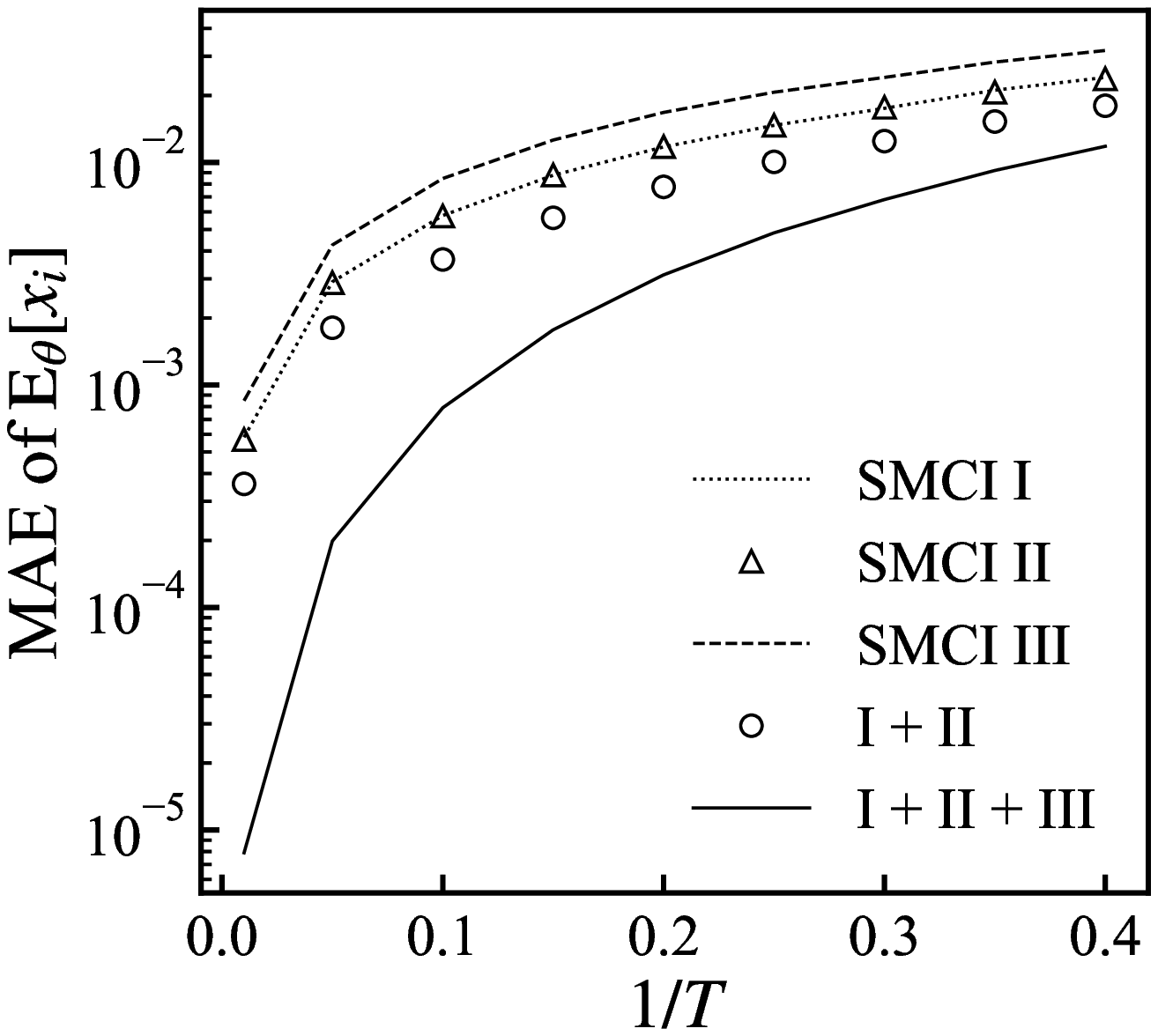} \\
\hspace*{0.86cm}(a)
\end{minipage}
\begin{minipage}[b]{0.45\linewidth}
\centering
\includegraphics[width=0.8\linewidth]{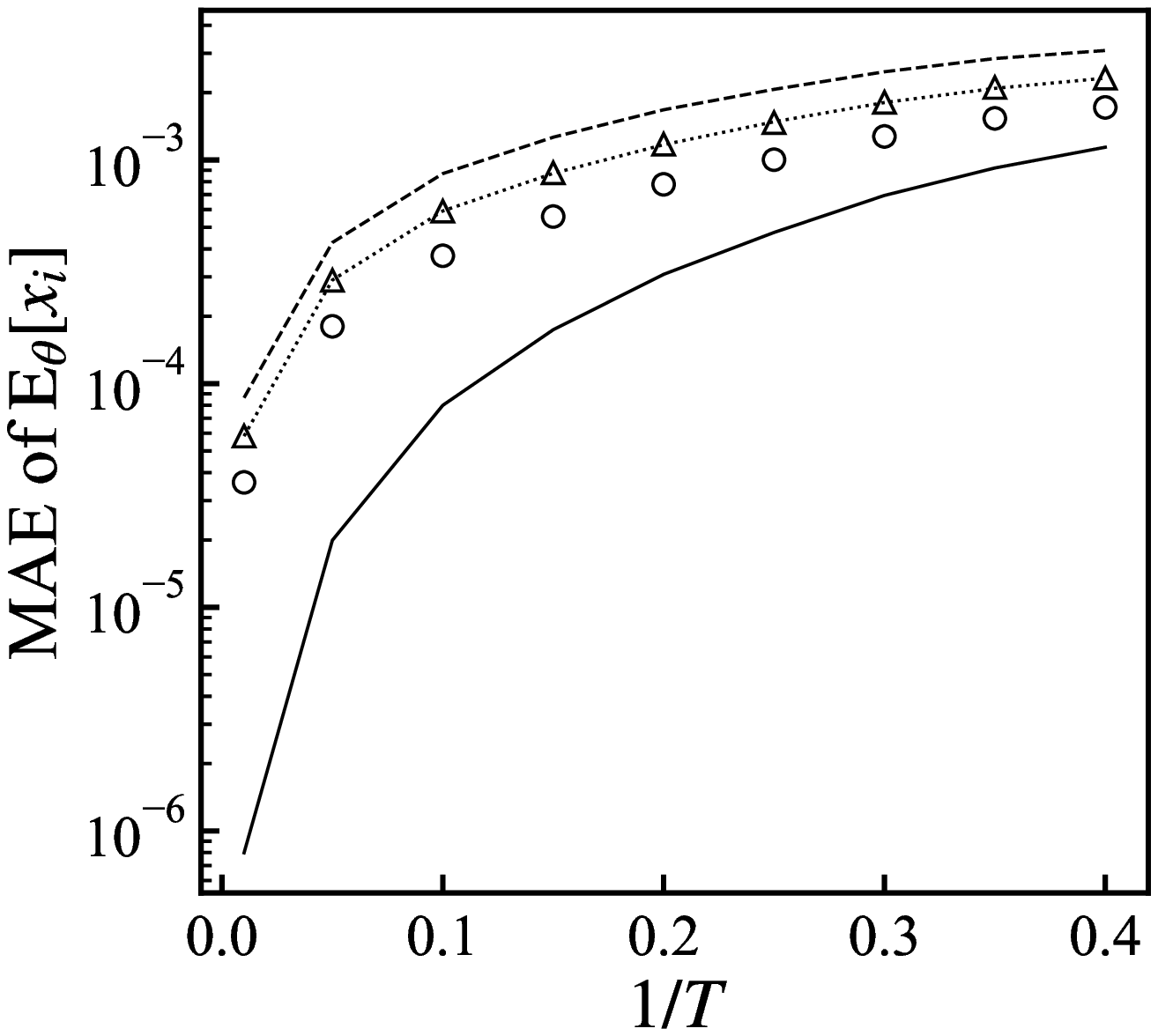} \\
\hspace*{0.9cm}(b)
\end{minipage}
\end{tabular}
\caption{MAEs of the evaluation of $\mexp{x_i}$ for $i \in \mcal{V}$ versus $1/T$ when (a) $N = 100$ and (b) $N = 10000$.
These plots are the average of $100$ experiments.}
\label{fig:MAE_versus_1/T_large}
\end{figure*}

Experiments on the Ising model defined on a $12\times12$ square-lattice graph with the free boundary condition are demonstrated, 
in which the external fields, $h_i$, are fixed to zero and the sample spaces are $\mcal{X}=\{-1,+1\}$. 
On this system, we again approximate the uni-variable expectation by the proposed estimator.
Obviously, $\mrm{E}_{\theta}[x_i]=0$; therefore, we can compute the MAE in equation (\ref{eq:MAE_E[xi]}).
The three different sum regions, $\mcal{U}_{\mrm{I}}$, $\mcal{U}_{\mrm{II}}$, and $\mcal{U}_{\mrm{III}}$, shown in Figure \ref{fig:sum_region_i} were again considered. 
Note that when the target region is at the edge of the graph, the overhang of the sum region was cut off (see Figure \ref{fig:sum_region_i_edge_case}).
The settings for $J_{i,j}$ and $r$ were the same as those in the experiments of Figure \ref{fig:MAE_versus_1/T}.
Figure \ref{fig:MAE_versus_1/T_large} shows the plots of the MAEs against $1/T$. 
The results in Figure \ref{fig:MAE_versus_1/T_large} are quantitatively almost the same as those in Figure \ref{fig:MAE_versus_1/T}.

\section{Application to the Inverse Ising Problem} \label{sec:inverse_ising_problem} 

\begin{figure*}[t]
\centering
\begin{tabular}{ccc}
\begin{minipage}[b]{0.45\linewidth}
\centering
\includegraphics[width=\linewidth]{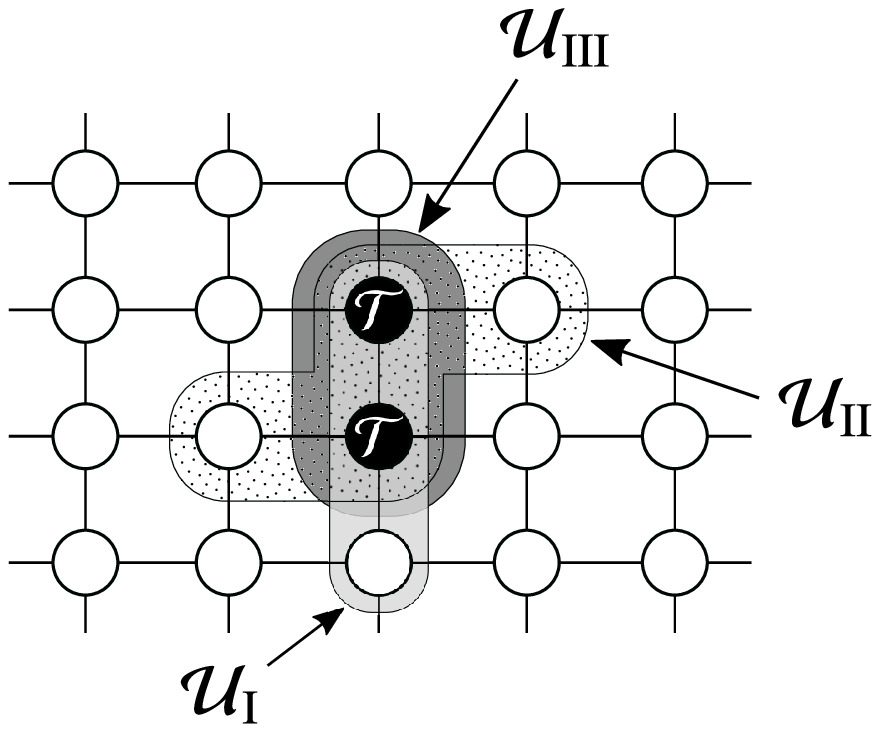} \\
(a)
\end{minipage}
\begin{minipage}[b]{0.45\linewidth}
\centering
\includegraphics[width=\linewidth]{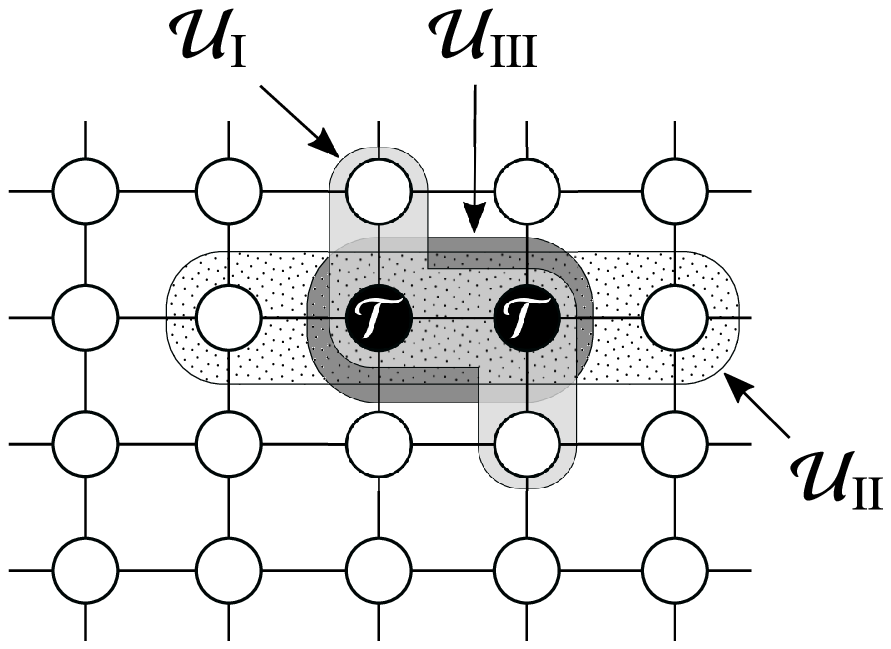} \\
(b)
\end{minipage}
\end{tabular}
\caption{Three different sum regions, namely, $\mcal{U}_{\mrm{I}}$, $\mcal{U}_{\mrm{II}}$, and $\mcal{U}_{\mrm{III}}$, for $\mcal{T}=\{i, j\}$: 
the cases of (a) $i$ and $j$ lining in a vertical direction and (b) $i$ and $j$ lining in a horizontal direction.}
\label{fig:sum_region_ij}
\end{figure*}

In this section, we apply the proposed method to the inverse Ising problem, which is also known as the Boltzmann machine learning in the machine-learning field.
Suppose that a set of $M$ i.i.d. data points: $\mcal{D} := \big\{\mbf{d}^{(\mu)}\mid \mu=1,2,\cdots,M\big\}$, 
where $\mbf{d}^{(\mu)} := \bigl\{\mrm{d}_i^{(\mu)}\in\mcal{X}\mid i\in\mcal{V}\bigr\}$ is the $\mu$th data point, is obtained. 
For the dataset, consider the log-likelihood that is defined by 
\begin{align}
\psi(\theta):= \frac{1}{M}\sum_{\mu=1}^M \ln P_{\theta}(\mbf{d}^{(\mu)} ).
\label{eq:log-likelihood}
\end{align} 
The inverse Ising problem is solved by the maximization of this log-likelihood with respect to $\theta$, namely, the ML estimation. 
The maximization is performed using a gradient ascent method. 
The gradients of the log-likelihood with respect to $h_i$ and $J_{i,j}$ are 
\begin{align}
\frac{\partial \psi(\theta)}{\partial h_i} &= \frac{1}{M}\sum_{\mu=1}^M \mrm{d}_i^{(\mu)} - \mexp{x_i}, 
\label{eq:gradient_of_h} \\
\frac{\partial \psi(\theta)}{\partial J_{i,j}} &= \frac{1}{M}\sum_{\mu=1}^M \mrm{d}_i^{(\mu)} \mrm{d}_j^{(\mu)}- \mexp{x_i x_j}.
\label{eq:gradient_of_J}
\end{align}
The first terms of these gradients are the sample averages of the dataset, and the second terms are the corresponding expectations of the Ising model.  
Because these gradients have the intractable expectations in their second terms, 
they have to be approximated to implement the ML estimation. 
We approximate these intractable expectations using the qCSMCI estimators proposed in section \ref{sec:CSMCI}, 
and examine the performance of the approximation through numerical experiments. 
In the experiments, two Ising models, defined on the same graph, were used. 
The first one is regarded as the generative model that generates the dataset, 
and the other one is regarded as the learning model that is used in the ML estimation. 
In the following experiments, the sample spaces of both generative and learning models were $\mathcal{X} = \{-1,+1\}$, and  
the learning rate (i.e., the step rate in the gradient ascent method) was fixed to $0.02$. 
The dataset, the size of which was fixed to $M=1000$, was obtained from the generative Ising model using Gibbs sampling, in which the number of the MCMC step, in both burn-in time and sampling interval, was fixed to $50$. 
The parameters of the learning model were initialized by zero.
To approximate the intractable expectations in equations (\ref{eq:gradient_of_h}) and (\ref{eq:gradient_of_J}) based on the SMCI and qCSMCI estimators, 
the sample set, consisting of $N$ sample point, generated from the learning model are required.  
To obtain the sample set, we used an $N$ parallel Gibbs sampling procedure that is similar to persistent contrastive divergence~\cite{tieleman2008}. 
In the $N$ parallel Gibbs sampling procedure, $N$ different sample points were initialized at random, 
because the initial state of the learning model is identified as uniform distribution. 
Next, the sample points were generated by performing the $N$ parallel Gibbs sampling procedure with the sampling interval of $\kappa$, 
starting from the current $N$ sample points,  
on the learning model updated by using the current $N$ sample points.

\begin{figure*}[t]
\centering
\begin{tabular}{cc}
\begin{minipage}[b]{0.45\linewidth}
\centering
\includegraphics[width=0.8\linewidth]{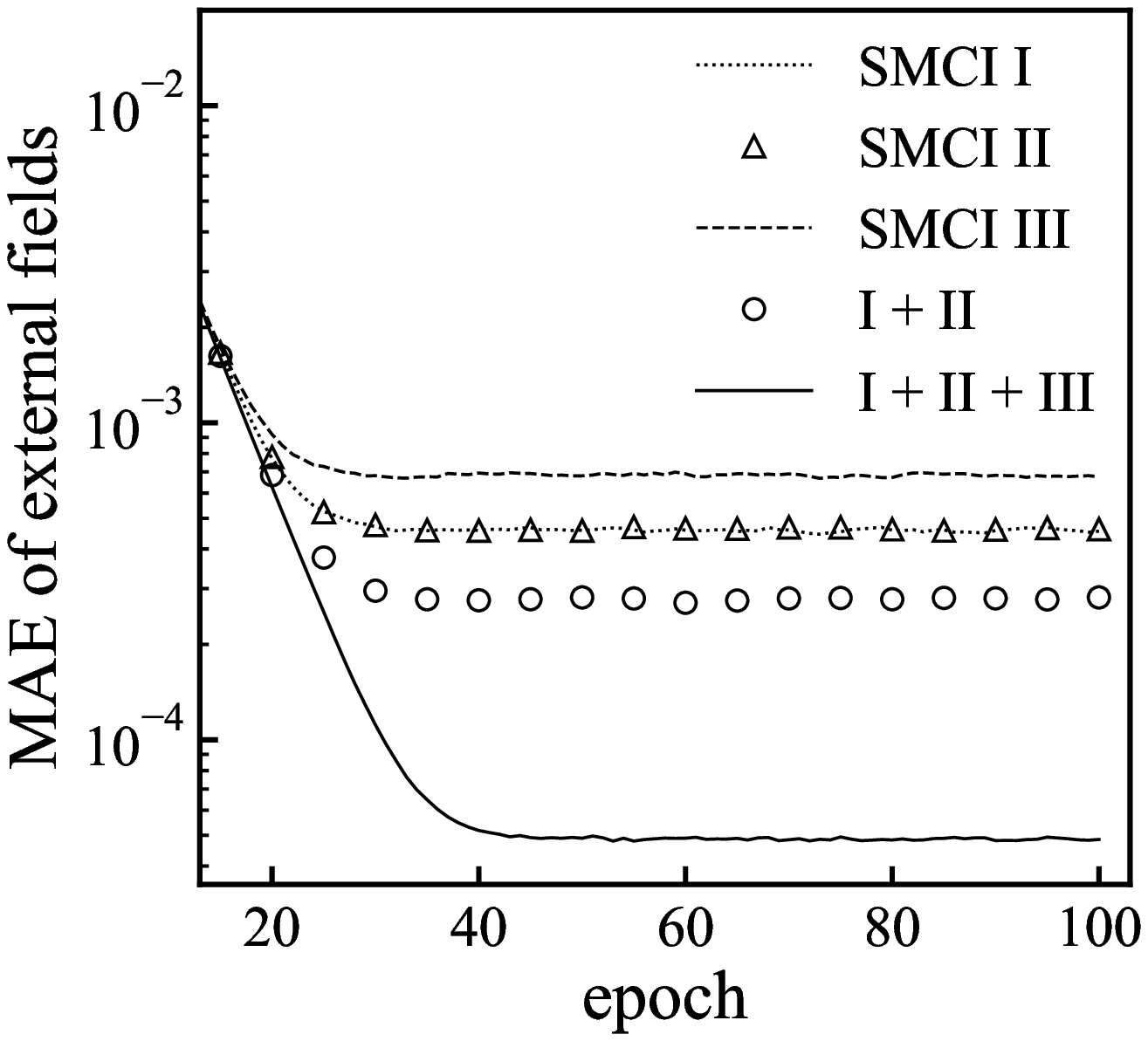} \\
\hspace*{0.85cm}(a)
\end{minipage}
\begin{minipage}[b]{0.45\linewidth}
\centering
\includegraphics[width=0.8\linewidth]{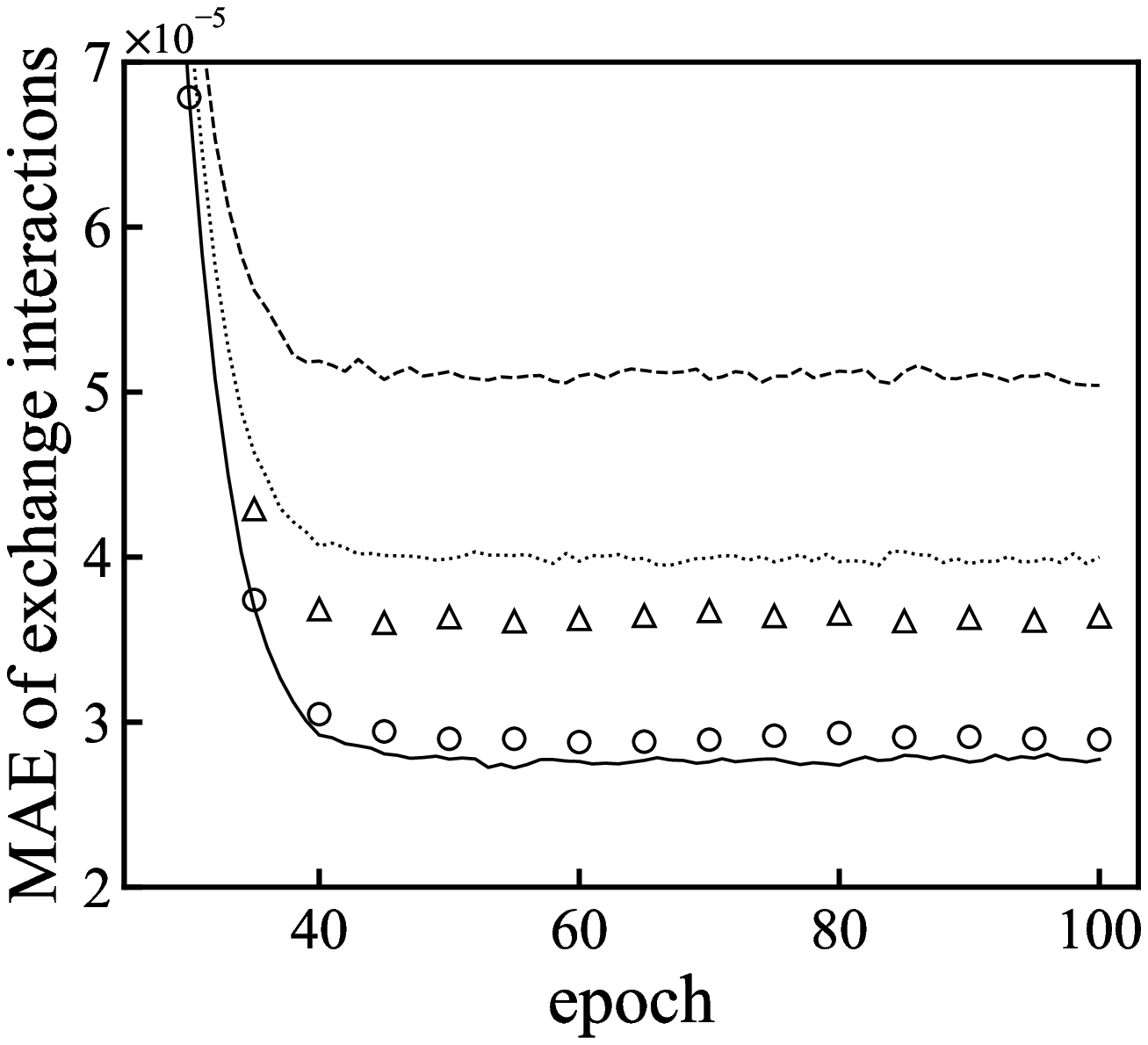} \\
\hspace*{0.5cm}(b)
\end{minipage}
\end{tabular}
\caption{MAEs of (a) $h_i$s for $i\in\mathcal{V}$ and (b) $J_{i,j}$s for $(i,j)\in\mathcal{E}$ versus the learning epoch when $1/T=0.05$. 
These results are the average of $500$ experiments.}
\label{fig:versus_epoch_when_1/T=0.05}
\end{figure*}

\begin{figure*}[t]
\centering
\begin{tabular}{cc}
\begin{minipage}[b]{0.45\linewidth}
\centering
\includegraphics[width=0.8\linewidth]{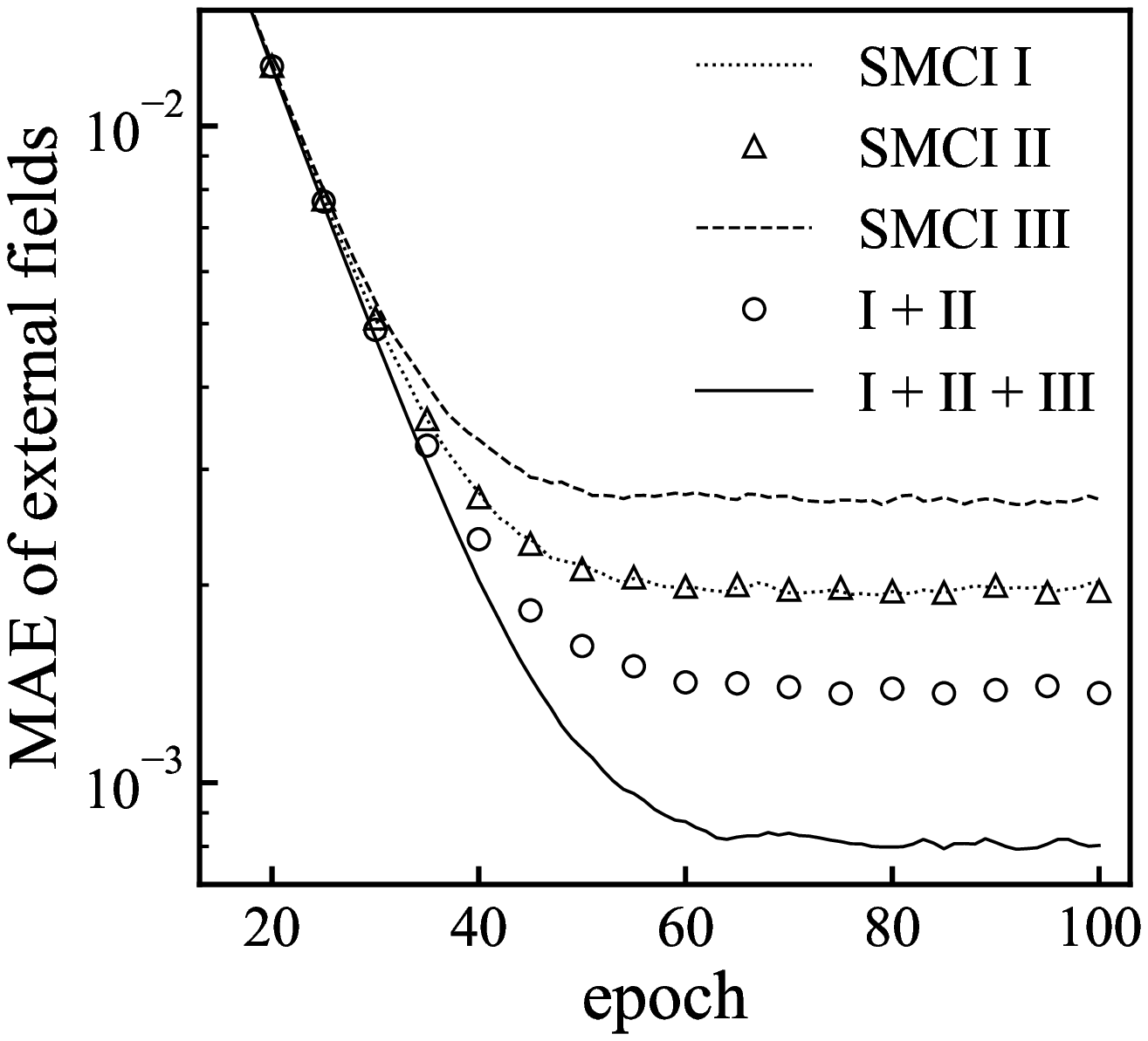} \\
\hspace*{0.85cm}(a)
\end{minipage}
\begin{minipage}[b]{0.45\linewidth}
\centering
\includegraphics[width=0.8\linewidth]{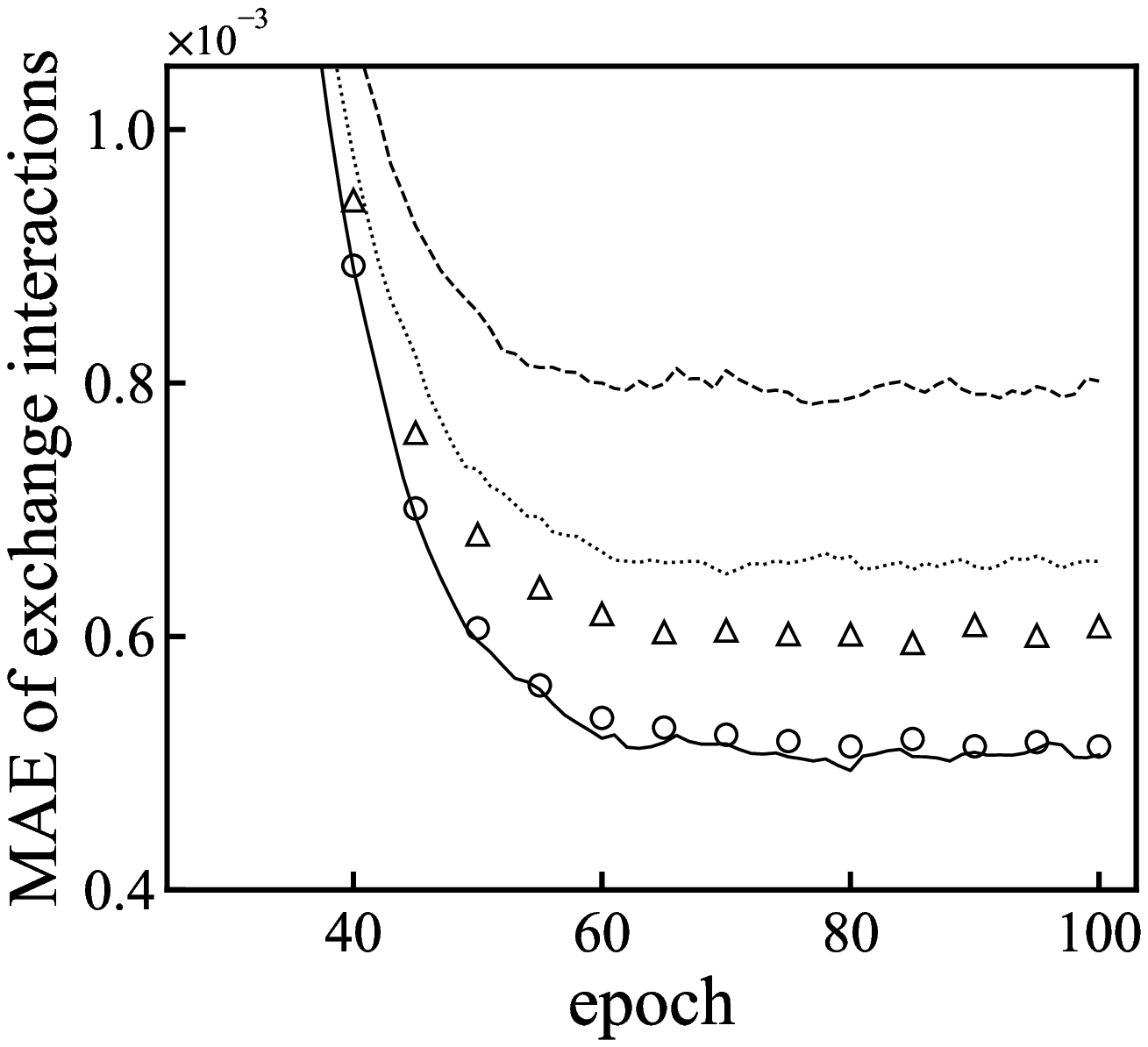} \\
\hspace*{0.7cm}(b)
\end{minipage}
\end{tabular}
\caption{MAEs of (a) $h_i$s for $i\in\mathcal{V}$ and (b) $J_{i,j}$s for $(i,j)\in\mathcal{E}$ versus the learning epoch when $1/T=0.3$. 
These results are the average of $500$ experiments.}
\label{fig:versus_epoch_when_1/T=0.3}
\end{figure*}

\begin{figure*}[t]
\centering
\begin{tabular}{cc}
\begin{minipage}[b]{0.45\linewidth}
\centering
\includegraphics[width=0.8\linewidth]{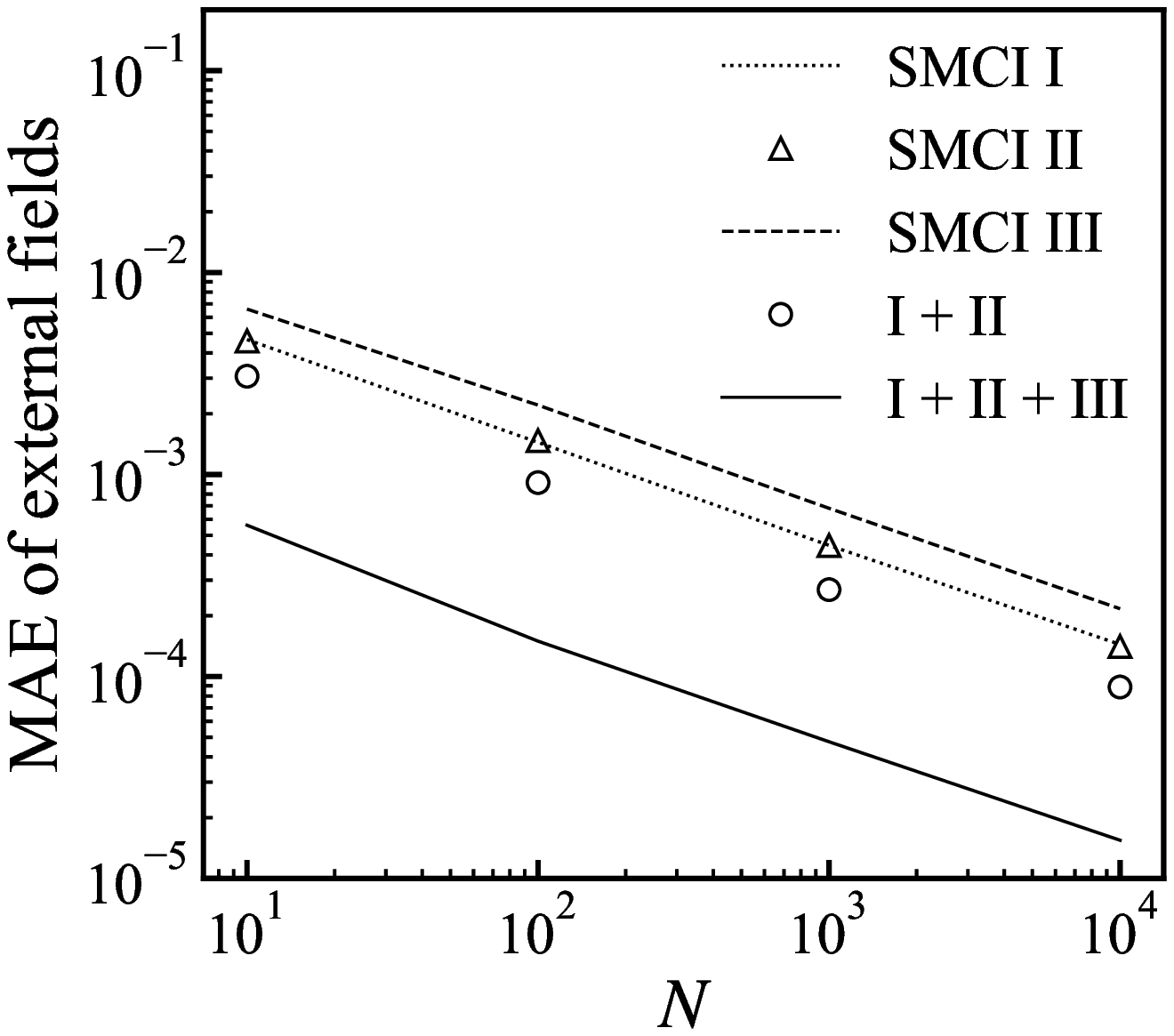} \\
\hspace*{0.89cm}(a)
\end{minipage}
\begin{minipage}[b]{0.45\linewidth}
\centering
\includegraphics[width=0.8\linewidth]{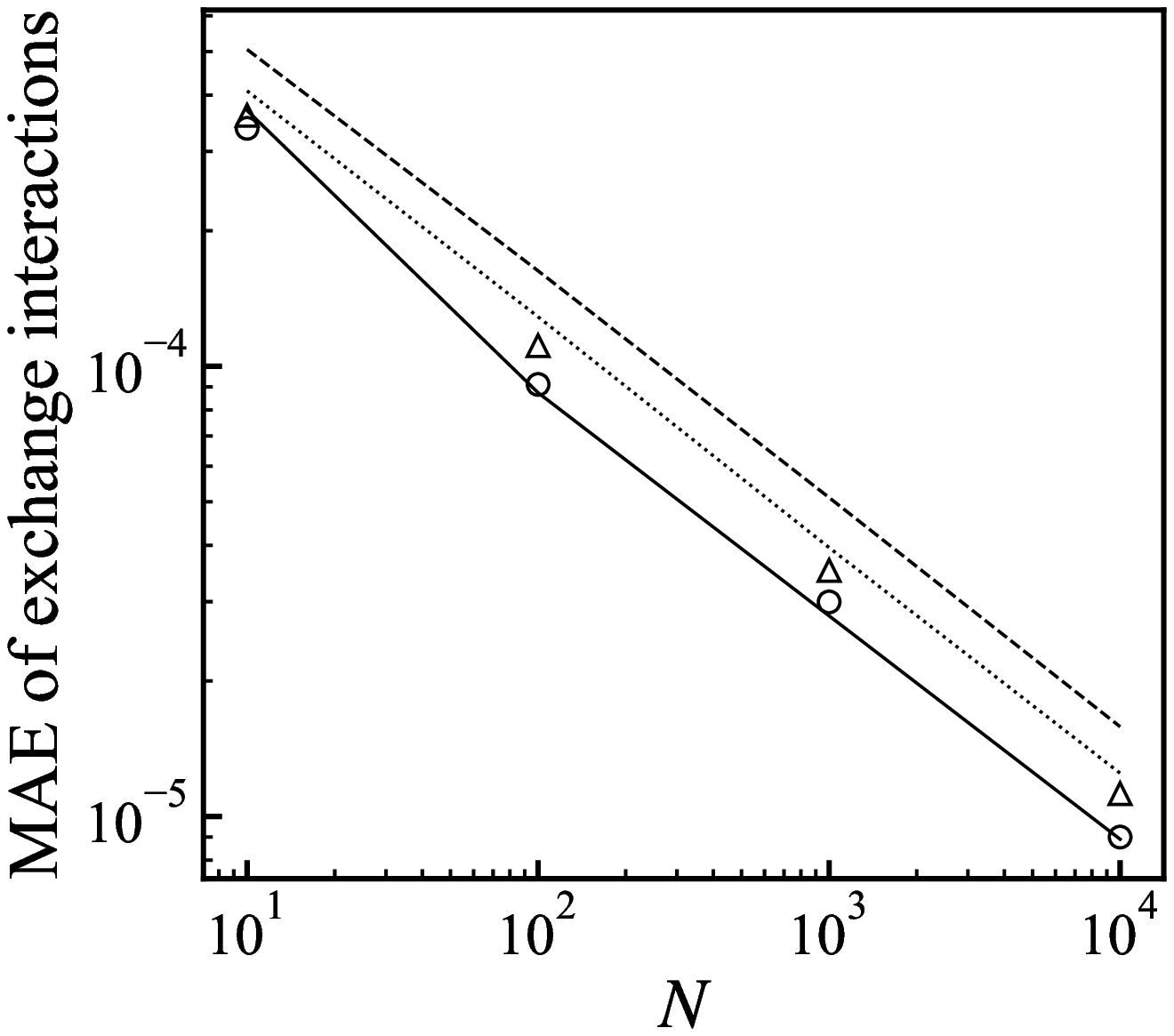} \\
\hspace*{0.89cm}(b)
\end{minipage}
\end{tabular}
\caption{MAEs of (a) $h_i$s for $i\in\mathcal{V}$ and (b) $J_{i,j}$s for $(i,j)\in\mathcal{E}$ versus $N$ when $1/T=0.05$. 
These results are the average of $500$ experiments.}
\label{fig:learning-MAE_versus_N_when_1/T=0.05}
\end{figure*}

\begin{figure*}[t]
\centering
\begin{tabular}{cc}
\begin{minipage}[b]{0.45\linewidth}
\centering
\includegraphics[width=0.8\linewidth]{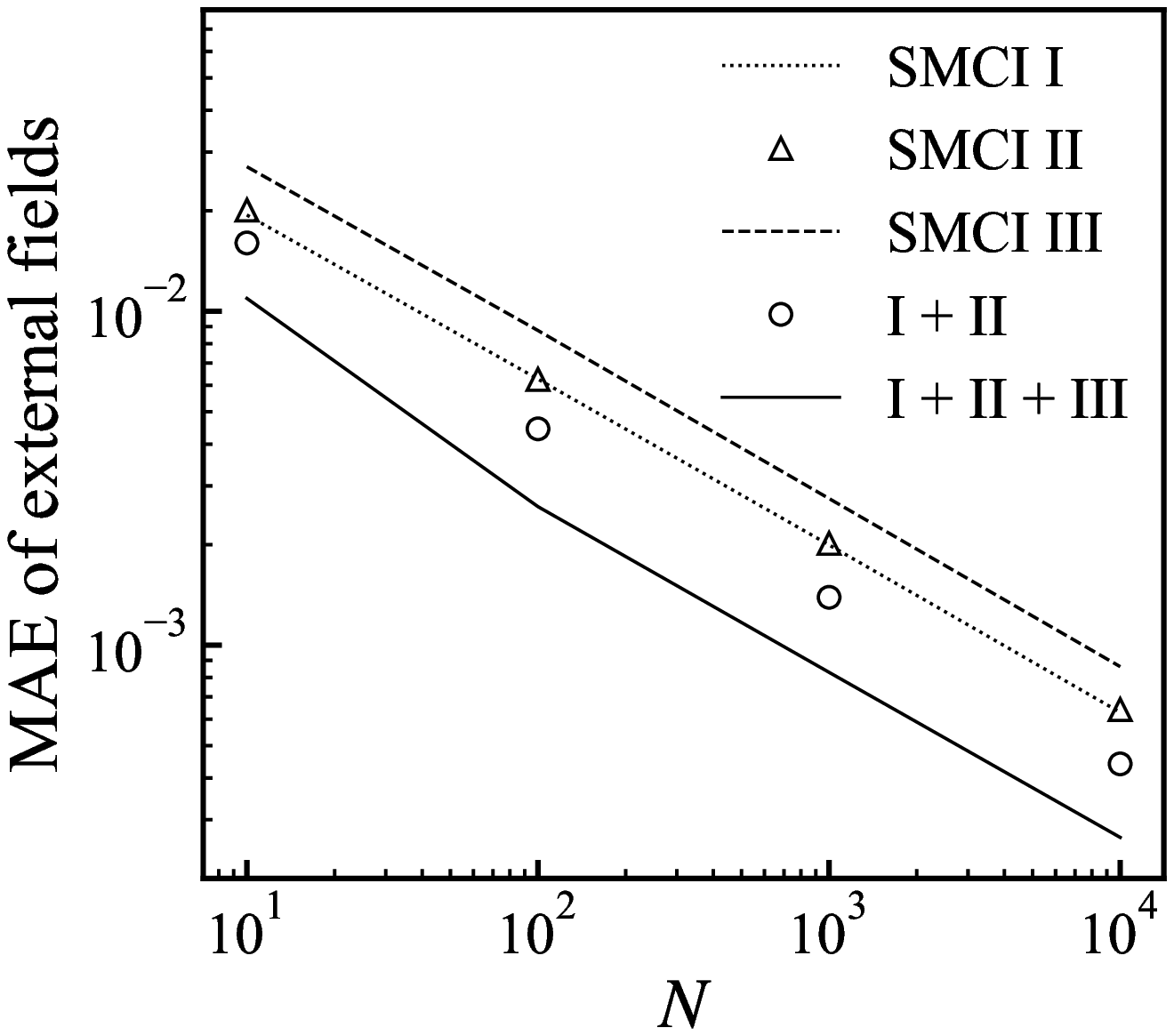} \\
\hspace*{0.89cm}(a)
\end{minipage}
\begin{minipage}[b]{0.45\linewidth}
\centering
\includegraphics[width=0.8\linewidth]{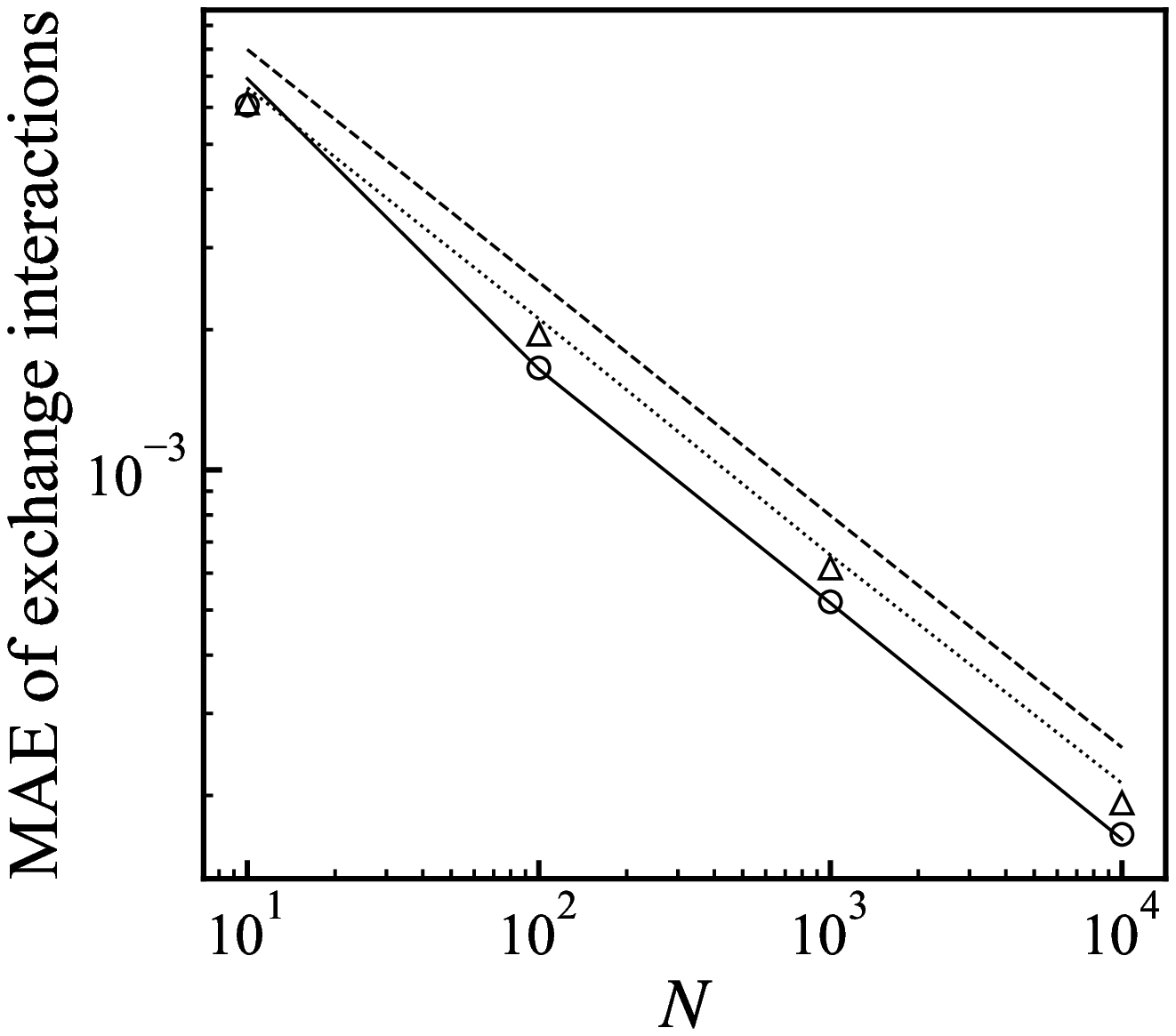} \\
\hspace*{0.89cm}(b)
\end{minipage}
\end{tabular}
\caption{MAEs of (a) $h_i$s for $i\in\mathcal{V}$ and (b) $J_{i,j}$s for $(i,j)\in\mathcal{E}$ versus $N$ when $1/T=0.3$. 
These results are the average of $500$ experiments.}
\label{fig:learning-MAE_versus_N_when_1/T=0.3}
\end{figure*}

\begin{figure*}[t]
\centering
\begin{tabular}{cc}
\begin{minipage}[b]{0.45\linewidth}
\centering
\includegraphics[width=0.8\linewidth]{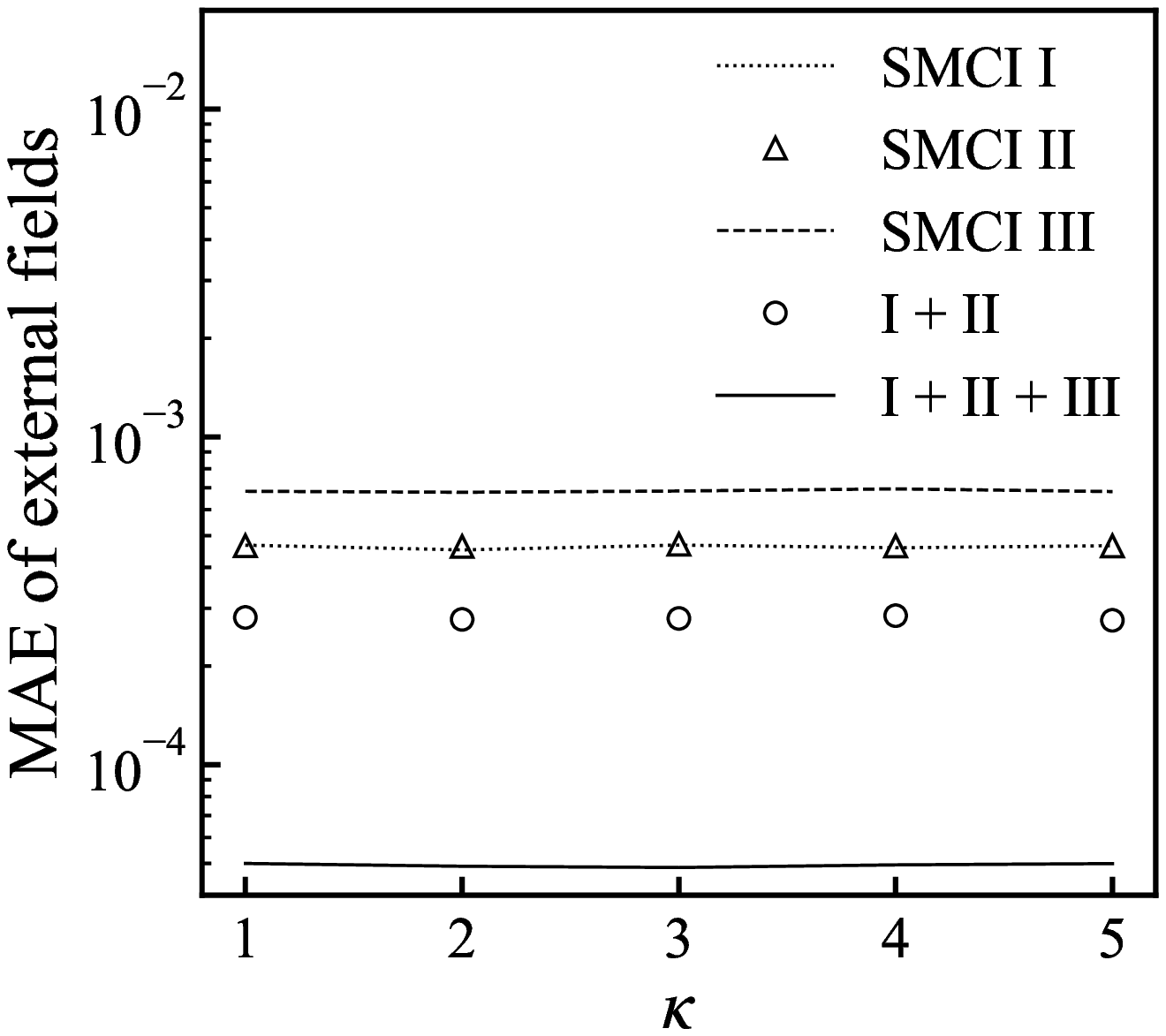} \\
\hspace*{0.89cm}(a)
\end{minipage}
\begin{minipage}[b]{0.45\linewidth}
\centering
\includegraphics[width=0.8\linewidth]{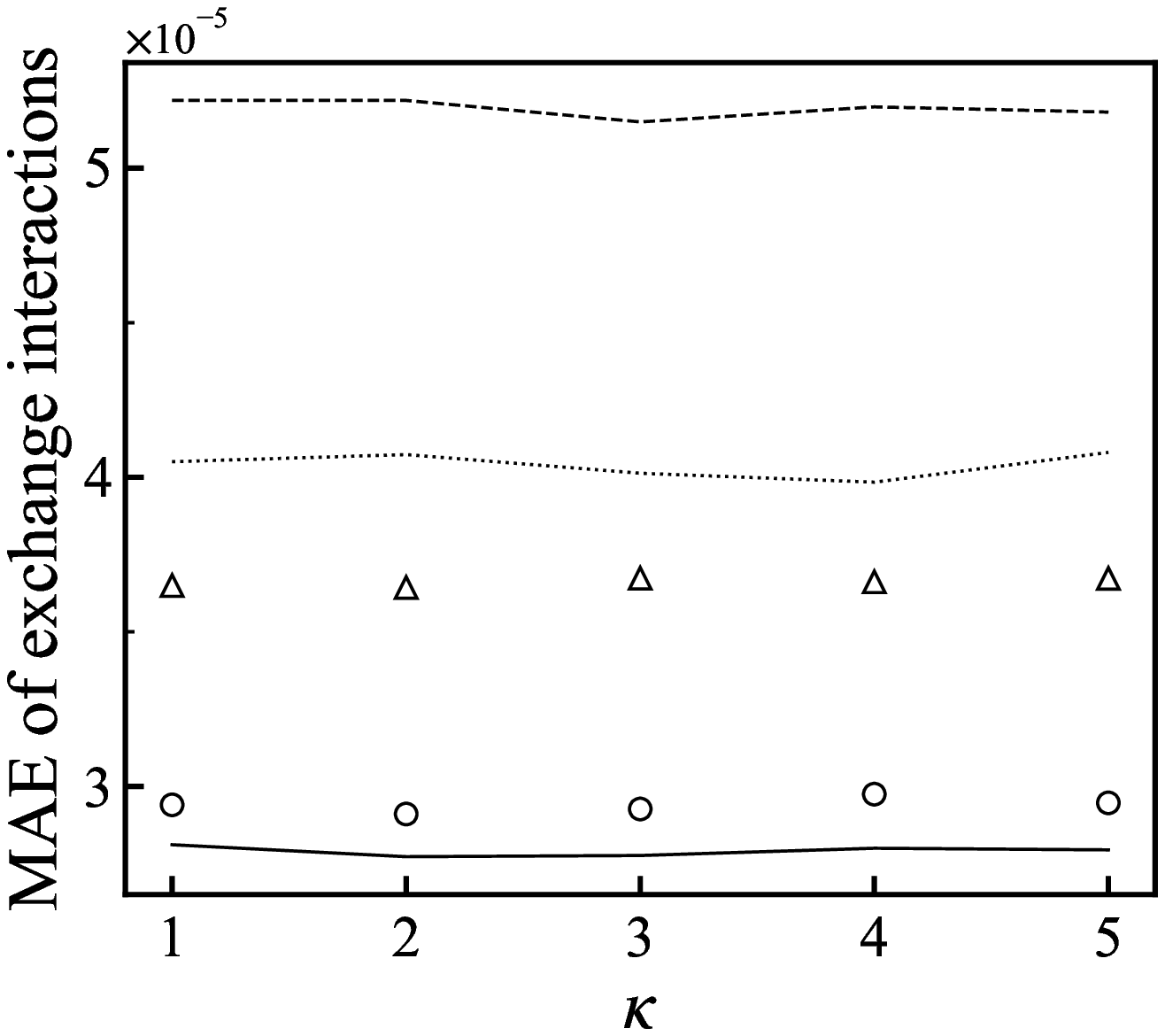} \\
\hspace*{0.6cm}(b)
\end{minipage}
\end{tabular}
\caption{MAEs of (a) $h_i$s for $i\in\mathcal{V}$ and (b) $J_{i,j}$s for $(i,j)\in\mathcal{E}$ versus $\kappa$ when $1/T=0.05$. 
These results are the average of $100$ experiments.}
\label{fig:learning-MAE_versus_MCMCstep_when_1/T=0.05}
\end{figure*}

\begin{figure*}[t]
\centering
\begin{tabular}{cc}
\begin{minipage}[b]{0.45\linewidth}
\centering
\includegraphics[width=0.8\linewidth]{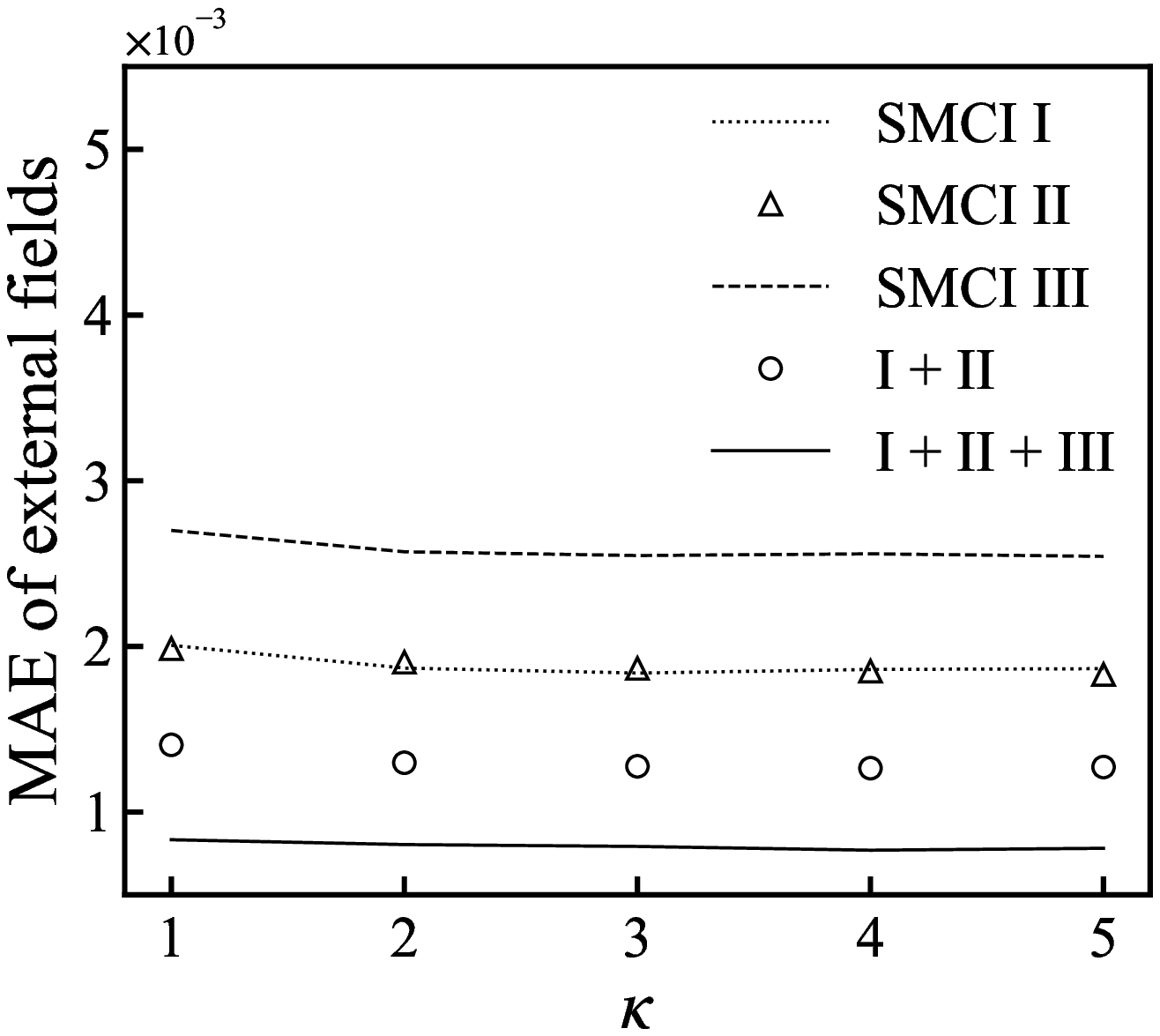} \\
\hspace*{0.55cm}(a)
\end{minipage}
\begin{minipage}[b]{0.45\linewidth}
\centering
\includegraphics[width=0.8\linewidth]{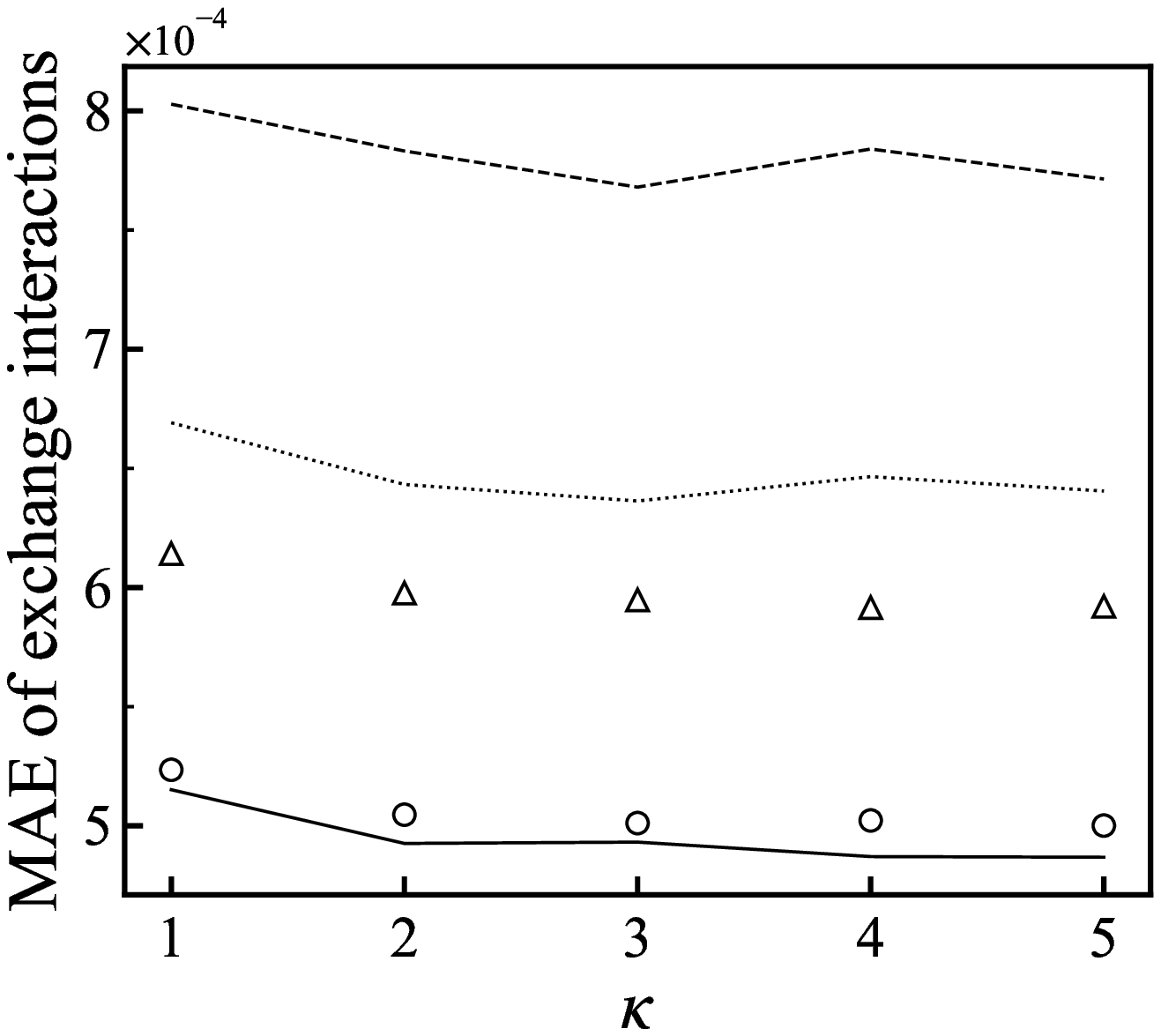} \\
\hspace*{0.6cm}(b)
\end{minipage}
\end{tabular}
\caption{MAEs of (a) $h_i$s for $i\in\mathcal{V}$ and (b) $J_{i,j}$s for $(i,j)\in\mathcal{E}$ versus $\kappa$ when $1/T=0.3$. 
These results are the average of $100$ experiments.}
\label{fig:learning-MAE_versus_MCMCstep_when_1/T=0.3}
\end{figure*}

First, we demonstrate the experiments in which the graph of both generative and learning models was a torus graph with $n = 20$ used in the experiments in section \ref{sssec:experiment_small}.
The parameters, $h_i$ and $J_{i,j}$, of the generative model were independently drawn from an uniform distribution in the interval $[-1/T, +1/T]$. 
Because $n$ is not large, we can obtain the exact ML estimations in this case. 
For the approximation of $\mexp{x_i}$ in equation (\ref{eq:gradient_of_h}), the three different sum regions shown in Figure \ref{fig:sum_region_i}, that is the same setting in the experiments in section \ref{sssec:experiment_small}, were used;  
and for $\mexp{x_ix_j}$ in equation (\ref{eq:gradient_of_J}), three different sum regions shown in Figure \ref{fig:sum_region_ij} were used.
The learning based on the SMCI estimators with the sum regions, $\mcal{U}_{\mrm{I}}$, $\mcal{U}_{\mrm{II}}$, and $\mcal{U}_{\mrm{III}}$, are referred to as ``SMCI-I,'' ``SMCI-II,'' and ``SMCI-III,'' respectively in the same manner as section \ref{sssec:experiment_small}. 
The SMCI-I corresponds to the 1-SMCI learning method, and 
the SMCI-II and SMCI-III corresponds to the semi-second-order learning method (without variable selection based on a greedy maximum independent set) proposed in the literature~\cite{yasuda2021}.
To compare with the SMCI learning methods, we considered two different qCSMCI learning methods. 
The first one is using SMCI-I and SMCI-II (referred to as  ``qCSMCI-I+II''), and the other one is using all three SMCI estimators (referred to as  ``qCSMCI-all'').  
The accuracy of learning was measured by the MAEs of the parameters compared with the those obtained from the exact ML estimation. 
That is, $n^{-1}\sum_{i\in\mcal{V}} \big| h_i^{(t)} - h_{i}^{\mrm{ML}} \big|$ 
and $|\mcal{E}|^{-1}\sum_{(i,j)\in\mcal{E}} \big|J_{i,j}^{(t)} - J_{i,j}^{\mrm{ML}}\big|$ were used, where $h_i^{(t)}$ and $J_{i,j}^{(t)}$ are the parameters obtained from the SMCI or qCSMCI method at $t$th epoch (or update) and $h_i^{\mathrm{ML}}$ and $J_{i,j}^{\mathrm{ML}}$ are the exact ML estimators.
 
Figures \ref{fig:versus_epoch_when_1/T=0.05} and \ref{fig:versus_epoch_when_1/T=0.3} show the MAEs against the number of the learning epochs (i.e., the number of the parameter updates). 
The numbers of the sample points and of the sampling interval were fixed to $N = M$ and $\kappa = 1$, respectively in these experiments.
In the figures, the results of ``I+II'' and ``I+II+III'' denote those obtained from qCSMCI-I+II and  qCSMCI-all, respectively. 
We observe that the proposed qCSMCI method improves the learning accuracy. 

Next, we investigated the dependency of the proposed method on $N$ and $\kappa$ values 
in which we used the learning models obtained after $t = 100$ epochs learning to evaluate the MAEs.
Figures \ref{fig:learning-MAE_versus_N_when_1/T=0.05} and \ref{fig:learning-MAE_versus_N_when_1/T=0.3} show the MAEs against $N$, in which $\kappa = 1$ was fixed. 
The MAEs decreased at a speed approximately proportional to $O(N^{-1/2})$. 
However, when $N$ is very small ($N \approx 10$), the MAEs of the exchange interactions of the proposed method are worse than those of the SMCI methods; 
the approximation error of the covariance matrix, $\bm{\Sigma}$, is not negligible in the case of small $N$. 
Figures \ref{fig:learning-MAE_versus_MCMCstep_when_1/T=0.05} and \ref{fig:learning-MAE_versus_MCMCstep_when_1/T=0.3} show the MAEs against 
the sampling interval $\kappa$ in the learning procedure, in which $N = M$ was fixed. 
A few $\kappa$ seems to be sufficient in the presented experiments.  
Because the magnitude of the learning rate used in these experiments was small,  
one update did not largely change the distribution of the learning model; 
therefore, the samplings were quickly relaxed. 
An appropriate value of $\kappa$ is strongly related to the magnitude of the learning rate. 
Alternatively, when the distribution of the learning model is strongly multimodal, say, it has a strong spin-glass-like property, 
a larger $\kappa$ would be required for the stable learning.

\begin{figure*}[t]
\centering
\begin{tabular}{cc}
\begin{minipage}[b]{0.45\linewidth}
\centering
\includegraphics[width=\linewidth]{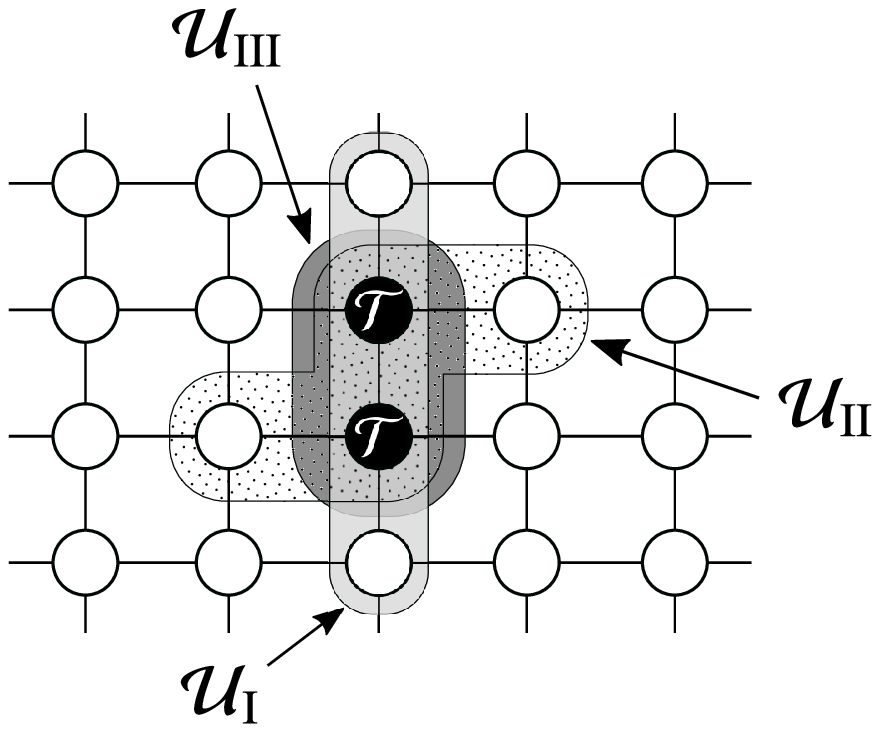} \\
(a)
\end{minipage}
\begin{minipage}[b]{0.45\linewidth}
\centering
\includegraphics[width=\linewidth]{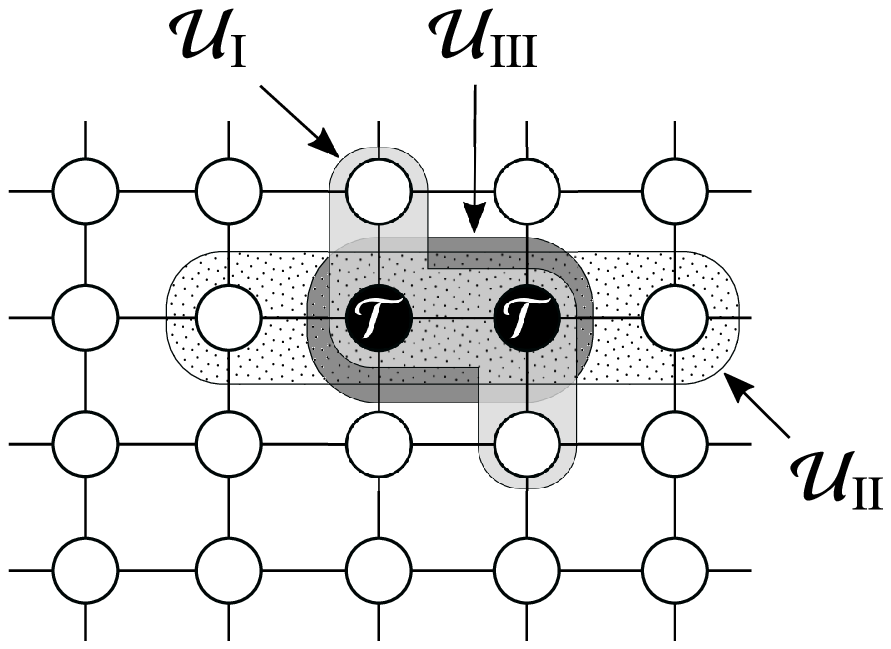} \\
(b)
\end{minipage}
\ \\
\ \\
\begin{minipage}[b]{0.45\linewidth}
\centering
\includegraphics[width=0.8\linewidth]{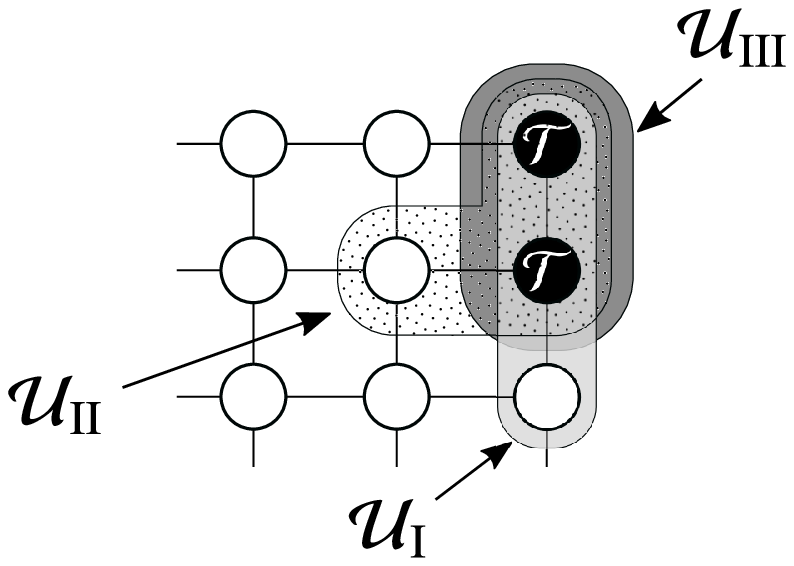} \\
(c)
\end{minipage}
\begin{minipage}[b]{0.45\linewidth}
\centering
\includegraphics[width=0.8\linewidth]{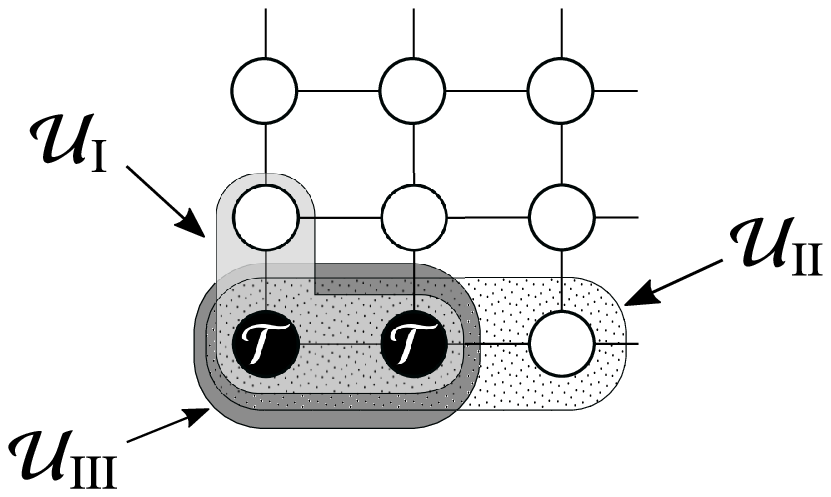} \\
(d)
\end{minipage}
\end{tabular}
\caption{Three different sum regions, $\mcal{U}_{\mrm{I}}$, $\mcal{U}_{\mrm{II}}$, and $\mcal{U}_{\mrm{III}}$, for $\mcal{T}=\{i, j\}$ on a $12\times12$ square lattice:  
the cases of (a) $i$ and $j$ lining in a vertical direction and (b) $i$ and $j$ lining in a horizontal direction, and the edge cases of (a) and (b) are (c) and (d), respectively.
In the edge cases, the overhangs of the regions are cut off.}
\label{fig:large_sum_region_ij}
\end{figure*}
\begin{figure*}[t]
\centering
\begin{tabular}{cc}
\begin{minipage}[b]{0.45\linewidth}
\centering
\includegraphics[width=0.8\linewidth]{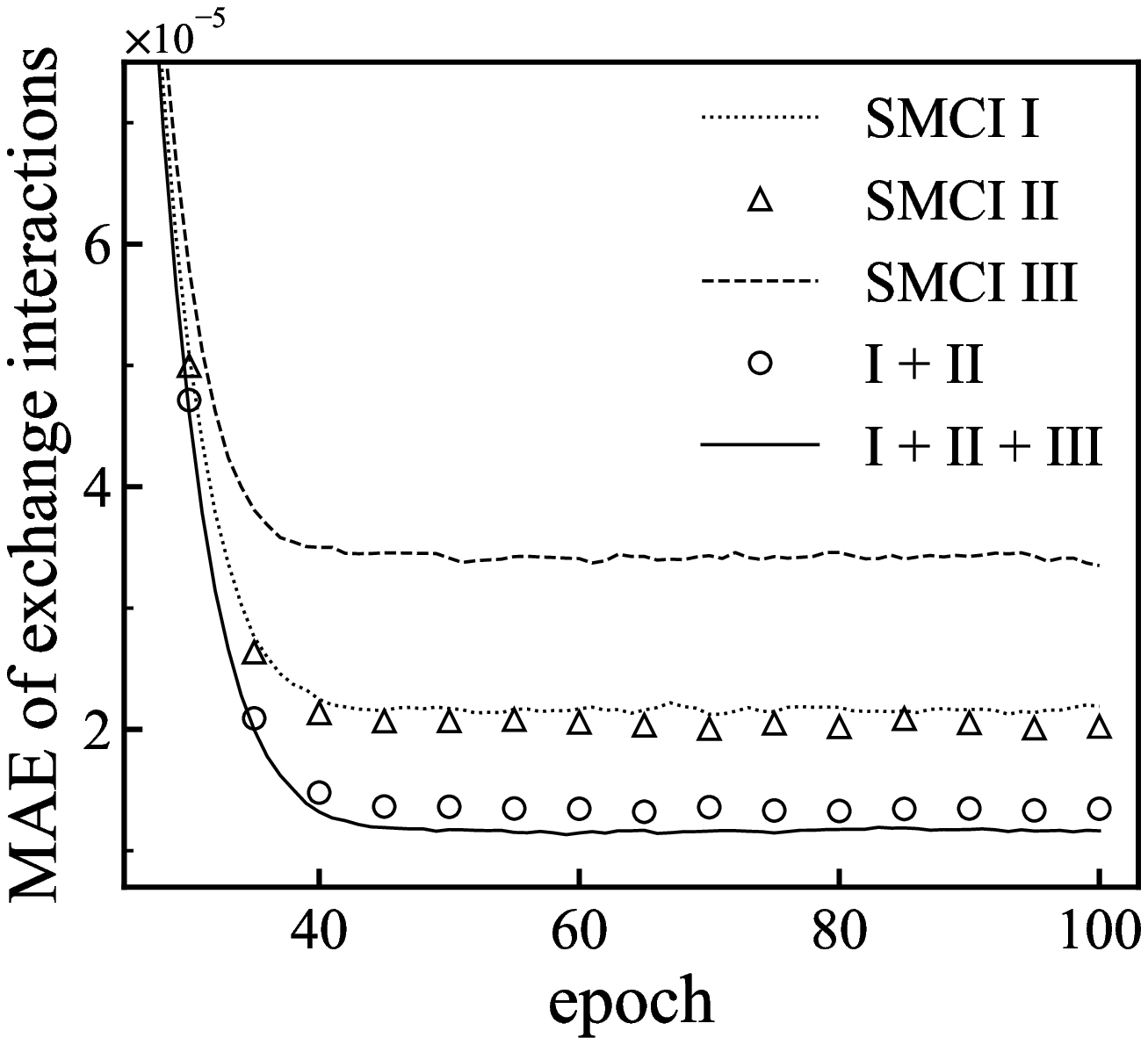} \\
\hspace*{0.89cm}(a)
\end{minipage}
\begin{minipage}[b]{0.45\linewidth}
\centering
\includegraphics[width=0.8\linewidth]{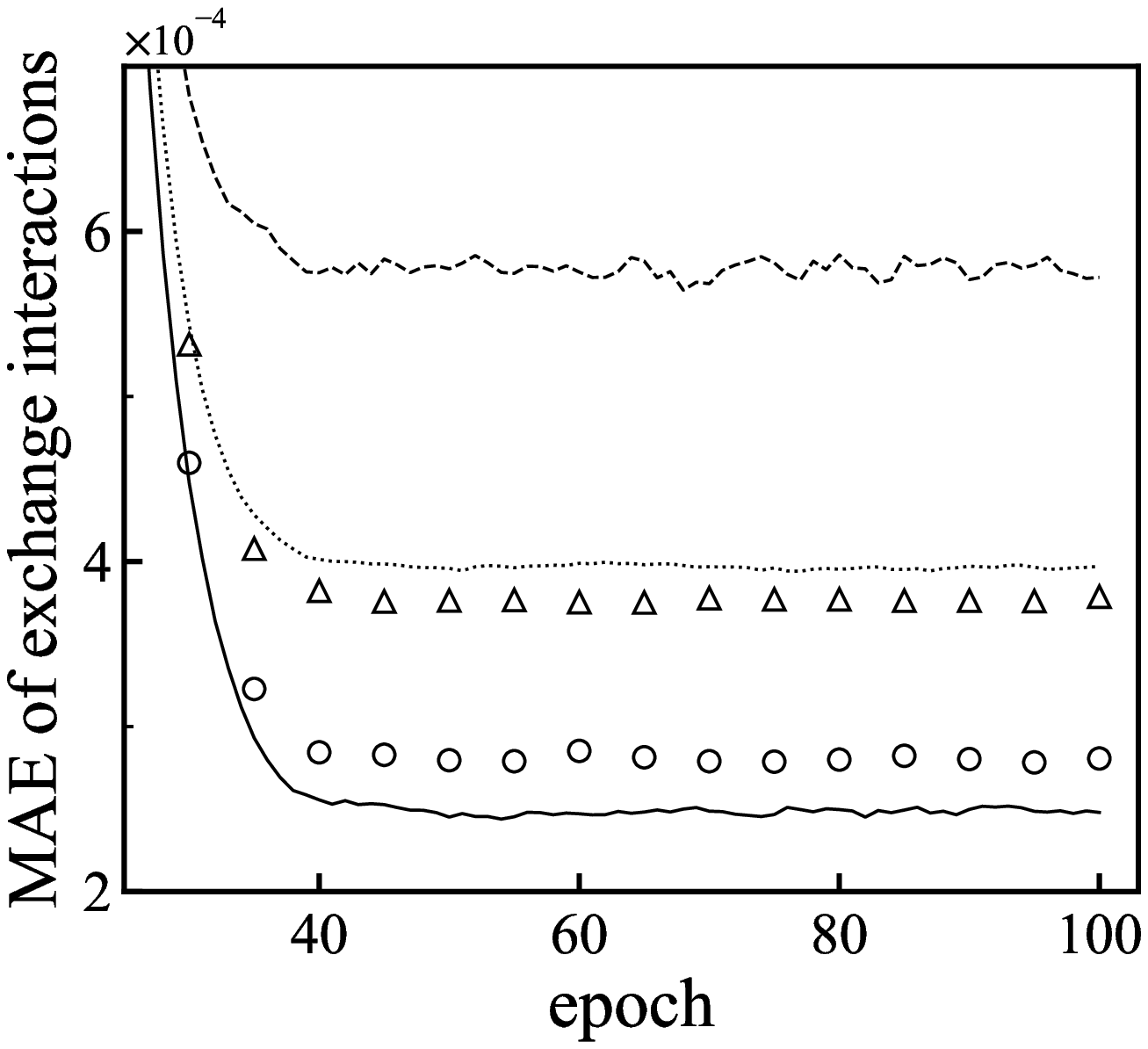} \\
\hspace*{0.89cm}(b)
\end{minipage}
\end{tabular}
\caption{MAEs of $J_{i,j}$s for $(i,j)\in\mathcal{E}$ versus the learning epoch when (a) $1/T=0.05$ and (b) $1/T=0.3$. 
These results are the average of $100$ experiments.}
\label{fig:learning-MAE_versus_epoch_large}
\end{figure*}

In the following, we demonstrate the experiments, in which the graph of both generative and learning models was a $12\times12$ square lattice used in the experiments in section \ref{sssec:experiment_large}.
The external fields of both generative and learning models were fixed at zero and the sample spaces of both models were $\mcal{X} = \{-1,+1\}$.
In this system (i.e., a square lattice without the external fields), 
we could obtain the exact ML estimations because $\mexp{x_i} = 0$ and $\mexp{x_ix_j}$ could be computed exactly~\cite{johnson2016}. 
For the approximation of $\mexp{x_ix_j}$, the three sum regions, $\mcal{U}_{\mrm{I}}$, $\mcal{U}_{\mrm{II}}$ and $\mcal{U}_{\mrm{III}}$, illustrated in Figure \ref{fig:large_sum_region_ij} were used. 
The interaction parameters of the generative model were independently drawn from an uniform distribution in the interval $[-1/T, +1/T]$.  
Figure \ref{fig:learning-MAE_versus_epoch_large} shows the MAEs against the learning epoch. 
The numbers of the sample points and the sampling interval were fixed to $N = M$ and $\kappa = 1$, respectively in these experiments.
This results look remarkably similar to those in Figures \ref{fig:versus_epoch_when_1/T=0.05} and \ref{fig:versus_epoch_when_1/T=0.3}.

\section{Summary and Future Works} \label{sec:conclusion} 

In this paper, we proposed a new estimator, called the CSMCI estimator, based on the theory of GLS, for an intractable expectation on the Ising model.
The CSMCI estimator was obtained as a linear combination of multiple SMCI estimators, $\bm{m}_{\mcal{T}}$, evaluating the same expectation 
(cf. equation (\ref{eq:GLS_estimator})). 
The proposed CSMCI estimator (without the covariance-matrix approximation) has a good property; that is, the CSMCI estimator is the BLUE, as well as the BUE; this means that it is the best among all possible unbiased estimators obtained from $\bm{m}_{\mcal{T}}$ (not necessarily linear). This leads to the following two properties: 
(1) the CSMCI estimator is guaranteed to improve the approximation accuracy 
and (2) the accuracy of the CMSCI estimator improves monotonically by adding a new SMCI estimator (cf. section \ref{ssec:theoretical}). 

However, the CSMCI estimator has intractable covariance matrix $\bm{\Sigma}$, which is the true covariance matrix among $\bm{m}_{\mcal{T}}$. 
Therefore, for the purpose of the implementation, we proposed the qCSMCI estimator 
that is obtained by replacing $\bm{\Sigma}$ with the sample covariance matrix (cf. equation (\ref{eq:CSMCI_estimator})). 
Unfortunately, these properties of the CSMCI estimator are not theoretically guaranteed in the qCSMCI estimator. 
However, as observed, the behavior of the qCSMCI estimator agrees with these properties of the CSMCI estimator 
in the numerical experiments in sections \ref{ssec:numerical} and \ref{sec:inverse_ising_problem}. 

The most important problem of the proposed method is the approximation of $\bm{\Sigma}$. 
As mentioned previously, we approximate it by the sample covariance matrix. 
However, a more appropriate approximation may exist. 
For example, an approach based on the ML estimation can be available. 
As discussed in section \ref{ssec:theoretical}, the framework of GLS can be regarded as the ML estimation for the log-likelihood of equation (\ref{eqn:log-likelihood_GLS}). 
A simultaneous maximization of the log-likelihood with respect to $\alpha$ and $\bm{\Sigma}$ will provide an alternative qCSMCI estimator. 
Finding a more effective approximation of $\bm{\Sigma}$ is addressed in our near-future research.
Additionally, applications of the proposed method to more practical learning models (e.g., RBMs and DBMs) are an important problem, will be addressed in our future project.

\subsection*{Acknowledgment}

This work was supported by JSPS KAKENHI (grant numbers: 18K11459, 18H03303, and 21K11778) and JEES / Softbank AI scholarship.

\appendix

\section{Approach Based on Lagrange Multipliers} \label{app:lagrange} 

The CSMCI estimator obtained in equation (\ref{eq:GLS_estimator}) can be obtained from an alternative strategy based on Lagrange multipliers.
Consider an unbiased estimator $\gamma_{\mcal{T}}$ for $\mexp{f(\bm{x}_{\mcal{T}})}$ 
expressed by a linear combination of the $K$ SMCI estimators, $\bm{m}_{\mcal{T}}$, as
$\gamma_{\mcal{T}} := \bm{w}^{\mrm{t}} \bm{m}_{\mcal{T}}$,
where $\bm{w} \in \mathbb{R}^K$ is the coefficient vector satisfying $\sum_{k=1}^K w_i = \bm{w}^{\mrm{t}} \bm{1}_K = 1$. 
Through this constraint, the unbiasedness of this estimator is ensured: $\sexp{\gamma_{\mcal{T}}} = \mexp{f(\bm{x}_{\mcal{T}})} $. 
The variance of the estimator is 
\begin{align}
\mrm{V}_{\mcal{S}}[\gamma_{\mcal{T}}] = \bm{w}^{\mrm{t}} \bm{\Sigma} \bm{w},
\label{eq:variance_gamma}
\end{align}
where $\bm{\Sigma}$ is the covariance matrix of $\bm{m}_{\mcal{T}}$ discussed in section \ref{sec:CSMCI}.

In the statistics perspective, the optimal $\bm{w}$ minimizes the variance in equation (\ref{eq:variance_gamma}). 
To find the optimal $\bm{w}$, we solve the optimization problem; that is, 
\begin{align*}
\min_{\bm{w}} \mrm{V}_{\mcal{S}}[\gamma_{\mcal{T}}] \quad \mrm{s.t.} \quad \bm{w}^{\mrm{t}} \bm{1}_K = 1.
\end{align*}
This optimization problem can be easily solved by using a Lagrange multiplier $\lambda$, i.e., we minimize the Lagrangian,  
\begin{align*}
L(\bm{w}, \lambda) :=  \mrm{V}_{\mcal{S}}[\gamma_{\mcal{T}}] - \lambda \big( \bm{w}^{\mrm{t}} \bm{1}_K  - 1\big), 
\end{align*}
with respect to $\bm{w}$. From the external conditions of the Lagrangian, 
the optimal $\bm{w}$ is obtained as the form of 
\begin{align*}
\bm{w} = \frac{1}{\Omega(\bm{\Sigma}^{-1})}\bm{\Sigma}^{-1}\bm{1}_K.
\end{align*}
The optimal $\bm{w}$ is equivalent to $\bm{c}$ defined in equation (\ref{eq:coefficient_c}); 
therefore, the optimal $\gamma_{\mcal{T}}$, in the perspective of the variance, is equivalent to the CSMCI estimator in equation (\ref{eq:GLS_estimator}).

\section{Asymptotic Property of CSMCI Estimator} \label{app:BUE}

For a sufficient large $N$, since the distribution of the error vector $\bm{\varepsilon}$ asymptotically converges to $\mcal{N}(\bm{\varepsilon} \mid \bm{0}_K, \bm{\Sigma})$, 
the distribution of $\bm{m}_{\mcal{T}}$ in equation (\ref{eq:reg_form}) asymptotically converges to $\mcal{N}(\bm{m}_{\mcal{T}} \mid \alpha \bm{1}_K, \bm{\Sigma})$. 
Therefore, in this case, the CSMCI estimator in equation (\ref{eq:GLS_estimator}) can be regarded as the ML estimator of the log-likelihood, 
\begin{align}
\ell(\alpha):=\ln \mcal{N}(\bm{m}_{\mcal{T}} \mid \alpha \bm{1}_K, \bm{\Sigma})
=-\frac{K}{2} \ln (2\pi) - \frac{1}{2} \ln \det \bm{\Sigma} - \frac{1}{2} \big( \bm{m}_{\mcal{T}} - \alpha \bm{1}_K\big)^{\mrm{t}} \bm{\Sigma}^{-1}
\big( \bm{m}_{\mcal{T}} - \alpha\bm{1}_K\big),
\label{eqn:log-likelihood_GLS}
\end{align}
with respect to $\alpha$, given $\bm{\Sigma}$.
From the log-likelihood, the Fisher information is obtained by
\begin{align*}
\SEXP{ \Big(\frac{\partial \ell(\alpha)}{\partial \alpha} \Big)^2}_{\alpha =m_{\mcal{T}}^{\mrm{true}} }
=\Omega(\bm{\Sigma}^{-1}). 
\end{align*}
From equation (\ref{eq:variance_GLS_estimator}), it is found that the inverse of the Fisher information  is equivalent to the variance of the CSMCI estimator,  
which implies that  the variance of the CSMCI estimator achieves with the  Cram\'er--Rao lower bound. 
From this, we conclude that the CSMCI estimator asymptotically is the BUE.

\bibliographystyle{jpsj}
\bibliography{reference}

\end{document}